\documentclass[12pt]{article}
\usepackage[margin=0.95 in]{geometry}
\usepackage{amsmath}
\usepackage{amssymb,amsfonts}
\usepackage[all]{xy}
\usepackage{graphicx}
\usepackage[utf8x]{inputenc}
\usepackage{amsmath}
\usepackage{amssymb}
\usepackage{float}
\usepackage{array}
\usepackage{tikz}
\usepackage{mathtools}
\usepackage{mathrsfs} 
\usepackage{hyperref}  
\usepackage{cite}
\numberwithin{equation}{section}
\numberwithin{equation}{section}
\numberwithin{table}{section}
\begin{document}

\thispagestyle{empty}

\vspace*{3cm}
{}

\noindent
{\LARGE \bf On the $gl(1|1)$ Wess-Zumino-Witten Model }
\vskip .4cm
\noindent
\linethickness{.06cm}
\line(10,0){447}
\vskip 1.1cm
\noindent
\noindent
{\large \bf Jan Troost}
\vskip 0.25cm
{\em 
\noindent
Laboratoire de Physique Th\'eorique 
de l'\'Ecole Normale Sup\'erieure \\
\hskip -.05cm
 CNRS,  
 PSL Research University and Sorbonne Universit\'es, Paris, France
}
\vskip 1.2cm

\vskip0cm

\noindent
{\sc Abstract: } {We continue the study of the $gl(1|1)$ Wess-Zumino-Witten
  model. The Knizhnik-Zamolodchikov equations for the one, two, three and four point
functions are analyzed, for vertex operators corresponding to  typical and projective
representations. We demonstrate their interplay with the  logarithmic
global conformal Ward identities.  
We  compute the  four point function
for one projective and three typical representations. Three
coupled first order Knizhnik-Zamolodchikov equations are integrated consecutively in terms of generalized hypergeometric functions, and we assemble the solutions into a local correlator.
Moreover, we prove crossing symmetry of the four point function
 of four typical representations at generic momenta.
Throughout, the map between 
the $gl(1|1)$ Wess-Zumino-Witten model and symplectic fermions is exploited and extended.}

\vskip 1cm

\pagebreak

 %
\newpage
\setcounter{tocdepth}{2}
\tableofcontents
\section{Introduction}
We  study a logarithmic conformal field theory with a supergroup symmetry, a supergroup Wess-Zumino-Witten model.
The simple example we focus on is the $gl(1|1)$ Wess-Zumino-Witten model. It has been studied in \cite{Rozansky:1992rx,Rozansky:1992td}
 where the four point functions of typical
representations were established and a check on locality   was performed. 
The supergeometry and the spectrum were clarified in \cite{Schomerus:2005bf}, while the
relation to symplectic fermions \cite{Gaberdiel:1998ps,Kausch:2000fu} has been made more manifest in \cite{Creutzig:2008an}. Many
other remarkable properties of the model have been obtained (for instance in \cite{Creutzig:2007jy,LeClair:2007aj}).

In this paper, we revisit the solution of the model, and take an  algebraic perspective on the calculation of the correlation
functions. We exploit the symmetries of the theory maximally, combine techniques
from the logarithmic conformal field theory literature for solving the global conformal Ward identities
 with those available for solving Wess-Zumino-Witten models,
and analyze the interplay between them.

The left regular representation of the $GL(1|1)$ group acting on itself gives rise
to typical representations as well as four-dimensional projective representations of the $gl(1|1)$ algebra.\footnote{The typical representations are also
projective, strictly speaking. We reserve the word projective for the four-dimensional representations, for ease of expression.}
  That 
suggests that these are the current algebra primaries arising in the spectrum
of the Wess-Zumino-Witten model  \cite{Schomerus:2005bf}. 
Thus, we take an approach in which we manifestly work with both  typical and projective representations.

The state of the art in the calculation of correlation functions of the
model uses representations of the vertex operators in terms of free
fields. In \cite{Schomerus:2005bf} a ghost system is used. In \cite{Creutzig:2008an}
a map is introduced between the $gl(1|1)$
Wess-Zumino-Witten degrees of freedom and two free bosons coupled to
symplectic fermions \cite{Kausch:2000fu} via an orbifold.  The latter free field
representation has the advantage of containing the fermionic zero
modes that are an important property of the model on the
supergroup.\footnote{See e.g. \cite{Flohr:2003tc} for a detailed discussion of fermionic zero modes in the $bc$ ghost system, in contrast to the
symplectic fermion model.}  Thus, we will revisit the symplectic fermion approach to the free field representation
of the correlators, and we will observe a close analogy between the differential constraints exploited by \cite{Gaberdiel:1998ps,Kausch:2000fu} to solve for 
the correlation functions of symplectic fermions, and the Knizhnik-Zamolodchikov equations \cite{Knizhnik:1984nr}.
This correspondence may be expected from the map between the models detailed in \cite{Creutzig:2008an} and from the relation of the two models through gauging 
the bosonic directions in the supergroup \cite{Creutzig:2011cu}.

An underlying goal of our analysis is to render the space-time
supergroup symmetry of the model more manifest. We may hope to  export the knowledge we gain into more complicated Wess-Zumino-Witten supergroup models,
relevant to string theory on $AdS_3 \times S^3$, such as the
$PSU(1,1|2)$ model \cite{Bershadsky:1999hk,Berkovits:1999im,Gotz:2006qp,Ashok:2009xx,Troost:2011fd,Gaberdiel:2011vf,Gerigk:2012cq}. Indeed, it should be noted that part of the motivation of the formalism that gives
rise to supergroup Wess-Zumino-Witten models in string theory is to render the space-time symmetry manifest \cite{Berkovits:1999im}, and in this paper we carry this motivation over into the
analysis of the world sheet model.

 In our manifestly covariant framework, we
 will find occasion to derive new technical results.
These include results on the number of invariants in various tensor product representations, and their explicit
identification, the check of crossing symmetry
of the four point functions of the typical representations at all
values of the external momenta, and four point
correlators for a projective top state and three typical insertions, in terms
of generalized hypergeometric functions. The latter are reminiscent of results on other field theories with supergroup symmetries \cite{Maassarani:1996jn}.
The results we derive also contain within them new results on symplectic fermion correlators.

Section \ref{symmetries} of our paper discusses symmetries and resulting Ward identities  in a more general context.
In section \ref{gl11wzw} we introduce the $gl(1|1)$ Wess-Zumino-Witten model on which we concentrate. The logarithmic conformal
Ward identities are solved in section \ref{log}, and in section
\ref{correlatorsKZ} we start studying the consequences of the Knizhnik-Zamolodchikov equations in more detail. 
Finally, in section \ref{four} we integrate the four point functions of four typical, or three typical and one projective representation.
In section \ref{local}, we combine the left- and right-moving correlators into local   four point functions. We offer concluding
remarks in section \ref{conclusions}. Representation theoretic results on the $gl(1|1)$ algebra are gathered in appendix \ref{reps},
while appendix \ref{symplecticfermions} reviews useful properties of the symplectic fermion conformal field theory.

\section{Symmetries and Identities}
\label{symmetries}
The supergroup $gl(1|1)$ Wess-Zumino-Witten model is a logarithmic conformal
field theory \cite{Rozansky:1992rx,Rozansky:1992td}. Therefore, it satisfies conformal Ward identities that
take into account the logarithmic nature of the spectrum \cite{Flohr:2001zs}. Moreover,
the energy-momentum tensor of the theory can be constructed using a
generalization of the Sugawara construction \cite{Rozansky:1992rx}. As a consequence, the
correlators also satisfy generalized Knizhnik-Zamolodchikov equations \cite{Knizhnik:1984nr}.
In this section, we review these constraints on correlators arising
from the symmetries of the theory in a general context. We add   remarks
on the algebraic interplay between the logarithmic aspects of the model, and the
global supergroup symmetry.
\subsection{Logarithmic Conformal Field Theory}
When we bring the chiral scaling operator $L_0$ in our $gl(1|1)$ model into standard Jordan form, it contains Jordan cells of rank
two. Thus we have a logarithmic conformal field theory \cite{Rozansky:1992rx}. 
We will have conformal group transformation laws for the 
logarithmic conformal primary fields $\phi_h$ that read \cite{Flohr:2001zs}
\begin{eqnarray}
\phi_h(z) &=& (\partial f)^h (1+ \log (\partial f) \hat{\delta}_h) \phi_h (f(z)) \, .
\label{transformation}
\end{eqnarray}
The operator $\hat{\delta}_h$ codes the relation between primaries in a given Jordan cell.
For example, when we have two logarithmic primary operators $\Psi_h$ and $\Phi_h$ in a non-trivial Jordan cell of generalized conformal dimension $h$,
then they can be chosen such that the operator $\hat{\delta}_h$ acts as
\begin{eqnarray}
\hat{\delta}_{h} \Psi_{h} = \Phi_{h} \, ,
\qquad
\hat{\delta}_{h} \Phi_{h} = 0  \, .
\end{eqnarray}
The logarithmic action of the Virasoro generators that follows from the transformation law (\ref{transformation}) is
\begin{eqnarray}
L_n \langle \phi_1(z_1) \dots \phi_n(z_n) \rangle
&=& \sum_i (z_i^{n+1} \partial_i + z_i^n (n+1) (h_i + \hat{\delta}_{h_i}))  \langle \phi_1(z_1) \dots \phi_n(z_n)  \rangle \, .
\label{Lnaction}
\end{eqnarray}
On the sphere, we have three globally defined chiral Ward identities, namely, for $n=-1,0,1$  the action (\ref{Lnaction}) of the
generators $L_n$ equals zero.

Before we move on, let us remark that results on correlation functions in logarithmic conformal
field theory are often valid for correlators that involve only fields that are logarithmic themselves, or
that cannot generate a logarithmic field through operator multiplication. Our correlators will fall outside the domain of validity of these general results,
since we will have correlators of pre-logarithmic fields \cite{Flohr:2001zs}.
These pre-logarithmic fields are not logarithmic themselves, but do generate logarithmic fields through
 their operator product.\footnote{Fields in typical $\widehat{gl(1|1)}$ primary current algebra representations will occasionally multiply into
fields in  projective representations and while the first class is of Jordan rank one,
the second is of Jordan rank two.}

We have briefly reviewed the consequences of chiral global conformal symmetry
in logarithmic models, and turn to the implications of the affine current
algebra symmetry.

\subsection{The Knizhnik-Zamolodchikov Equation}
We will examine a theory with a generalized Sugawara energy-momentum tensor \cite{Rozansky:1992rx}. The holomorphic
energy-momentum tensor component $T(z)$ is quadratic in the holomorphic currents $J^a$ that capture the affine
symmetry of the Wess-Zumino-Witten model, 
\begin{eqnarray}
T &=& \frac{1}{2} Q_{ab} J^a J^b \, , \label{TQ}
\end{eqnarray}
where $Q_{ab}$ is a constant set of coefficients to be determined in due course.
We define current algebra primaries $\phi$
through the equations
\begin{eqnarray}
J^a(z) \phi(w) & \approx &  \frac{t^a}{z-w} \phi(w) + \mbox{regular}  \label{currentprimary}
\end{eqnarray}
where $t^a$ are representation matrices associated to the representation of the global symmetry
group under which the current algebra primary $\phi$ transforms. Given these algebraic structures, we can derive the Knizhnik-Zamolodchikov equation
for the model as a consequence of the identity between the first mode of the energy momentum tensor $L_{-1}$
and the bi-linear combination of the first and zeroth mode of the current algebra \cite{Knizhnik:1984nr} (when acting on current algebra primaries).
The Knizhnik-Zamolodchikov equation on the current algebra primary correlators on the sphere reads
\begin{eqnarray}
(\partial_i - \sum_{j \neq i} \frac{ Q_{ab} t^a_i \otimes t^b_j}{z_i-z_j})
\langle \phi_1(z_1) \dots \phi_n(z_n) \rangle &=& 0 \, , \label{KZ}
\end{eqnarray}
where $\partial_i$ denotes the derivative with respect to the holomorphic coordinate $z_i$ of 
the $i$-th insertion, and the summation is over all other insertions, at the points $z_j$.

\subsection{Common Consequences}
Often, global conformal Ward identities and the Knizhnik-Zamolodchikov equation are exploited conjointly. The conformal symmetry
can be used to reduce the dependence of the four point functions to a dependence on one variable, while the Knizhnik-Zamolodchikov equations can then
be exploited to integrate that dependence. In this subsection,
we pause to observe the interplay between logarithmic conformal Ward identities and the Knizhnik-Zamolodchikov equation.
Firstly, we note that the Knizhnik-Zamolodchikov equation implies the global conformal Ward identity for translations.
This is seen by summing over the choice of insertion $i$ in the equation (\ref{KZ}).
More interestingly, we can multiply the Knizhnik-Zamolodchikov equation (\ref{KZ}) by $z_i$ and sum over the insertion points $i$,
to obtain the equation
\begin{eqnarray}
(\sum_i z_i \partial_i - \sum_{i < j} Q_{ab}  t^a_i \otimes t^b_j) \langle  \phi_1(z_1) \dots \phi_n(z_n)  \rangle &=& 0
\, .
\end{eqnarray}
The scaling Ward identity on the other hand reads
\begin{eqnarray}
(\sum_i z_i \partial_i + \frac{1}{2} Q_{ab} t^a_i t^b_i 
) \langle \phi_1(z_1) \dots \phi_n(z_n) \rangle &=& 0 \, .
\end{eqnarray}
We have used that $L_0$ acts as the zero mode of the energy-momentum tensor on the current algebra primaries. The action of this operator
captures both the conformal dimension $h_i$ {\em and} the logarithmic term $\hat{\delta}_{h_i}$.
Combining these two equations gives rise to the identity
\begin{eqnarray}
(\frac{1}{2} Q_{ab} \sum_i t^a_i \otimes \sum_j t^b_j 
) \langle  \phi_1(z_1) \dots \phi_n(z_n)  \rangle &=& 0 \, ,
\end{eqnarray}
which is also implied by the global symmetry of the model.
Indeed, the quadratic Casimir\footnote{Namely, the particular quadratic Casimir fixed by the energy-momentum tensor.}
 $C_2^\otimes$ in the tensor product representation of a correlator
equals zero on that correlator
\begin{eqnarray}
C^\otimes_2 
\langle  \phi_1(z_1) \dots \phi_n(z_n) \rangle &=& 0 \, .
\end{eqnarray}
For the supergroup conformal field theory this simple equation has interesting consequences.

\section{The $gl(1|1)$ Wess-Zumino-Witten Model}
\label{gl11wzw}
In this section, we focus on the $gl(1|1)$ Wess-Zumino-Witten model. It is a conformal non-linear sigma-model
on a supergroup with Lie algebra $gl(1|1)$. To specify the structure of the model, we review the algebra,
the current algebra and the energy-momentum tensor. We discuss their representations and a free field
realization.
\subsection{The $gl(1|1)$ Algebra}
The $gl(1|1)$ Lie algebra has four generators, of which two are bosonic, and two fermionic.\footnote{See also appendix \ref{reps} for more details.} We denote
them $E,N$ and $\Psi^\pm$ respectively. The non-trivial commutation relations are
\begin{eqnarray}
\{ \Psi^+, \Psi^- \} &=& E \, , 
\nonumber \\
{[} N , \Psi^\pm {]} &=& \pm \Psi^\pm \, .
\end{eqnarray}
The supertrace in the product of the defining representation gives
a non-degenerate bi-linear form $\kappa$ that satisfies
\begin{eqnarray}
\kappa(N,E) = \kappa(E,N) &=& 1 \, ,
\nonumber \\
\kappa(\Psi^+,\Psi^-) = - \kappa(\Psi^-,\Psi^+) &=& 1 \, .
\end{eqnarray}
A quadratic Casimir is $C_2(\alpha)=EN+NE + \Psi^- \Psi^+- \Psi^+ \Psi^- + \alpha E^2$ where $\alpha$ is any  constant
since the generator $E$ is central.
\subsection{The Current Algebra}
The Wess-Zumino-Witten model has a holomorphic current algebra $\widehat{gl(1|1)}$ generated by the conserved current components
$J^a(z)$.
For each generator in the algebra, we obtain a current $J^a$ with Laurent coefficients $J^a_n$, and we find that they satisfy the affine $\widehat{gl(1|1)}$
current algebra:\footnote{The normalization of the non-zero level is arbitrary for this current algebra.}
\begin{eqnarray}
{[} N_r , E_{s} {]} &=&  r  \delta_{r+s} \nonumber \\
{[} N_r , \Psi^\pm_{s} {]} &=& \pm \Psi^{\pm}_{r+s} \nonumber \\
{\{} \Psi^+_r , \Psi^-_{s} {\}} &=& E_{r+s}+ r  \delta_{r+s}  \, .
\end{eqnarray}
The corresponding current algebra operator product expansions are
\begin{eqnarray}
N(z) E(w) & \approx & \frac{1}{(z-w)^2}
\nonumber \\
N(z) \Psi^{\pm}(w) & \approx & \pm \frac{\Psi^{\pm}(w)}{z-w}
\nonumber \\
\Psi^{+}(z) \Psi^-(w) & \approx & \frac{1}{(z-w)^2} + \frac{E(w)}{z-w}
\, .
\end{eqnarray} 
\subsection{The Energy-momentum Tensor}
A holomorphic energy-momentum tensor component that ensures that the currents
are primaries of conformal dimension one is  \cite{Rozansky:1992rx}
\begin{eqnarray}
T(z) &=& \frac{1}{2} : (NE + EN  +   \Psi^- \Psi^+ - \Psi^+ \Psi^- + EE) :
\, .
\end{eqnarray}
This requirement also fixes the constant in front of the term proportional to $E^2$. The central charge of the model is zero.
The energy-momentum tensor is a bi-linear in the currents, and is  of  Sugawara type.
We equate the above energy-momentum tensor to the formula (\ref{TQ}), thus fixing the tensor
$Q_{ab}$ which features in the Knizhnik-Zamolodchikov equation (\ref{KZ}).
\subsection{The Representation Theory and the Spectrum}
The representation theory of the $gl(1|1)$ algebra is briefly reviewed in appendix \ref{reps}. We will mostly work with 
typical  and projective cover representations. The typical representations $T_{e,n}$ are two-dimensional, 
and have a highest weight state $| {\uparrow} \rangle$ and a lowest weight state $|{ \downarrow} \rangle$. They are
characterized by a non-zero eigenvalue $e$ for the generator $E$ and eigenvalues $n \pm 1/2$ for the other Cartan generator $N$.
The projective covers $P_n$ are four-dimensional representations that arise at eigenvalue $e=0$ for the generator $E$.
There are four states, denoted $t,l,r,b$  (for top, left, right and bottom) with eigenvalues $n,n+1,n-1,n$ respectively for the
generator $N$. See appendix \ref{reps} for more  details.

The mini-superspace analysis of \cite{Schomerus:2005bf} shows that the spectrum of quadratically integrable functions on the
non-compact version of the supergroup decomposes with respect to the left regular action of the $GL(1|1)$ group on itself as
\begin{eqnarray}
L^2 (GL(1|1)) &=& \int_{e \neq 0} \, de \, dn T_{e,n}^L \otimes T_{-e,-n}^R \oplus \int \, dn P_n^L \, .
\end{eqnarray}
In the typical sector, the left and right actions factorize. In the  projective ($e=0$) sector, the left projective representations $P_n^L$ combine into
infinite dimensional representations $J_q$ of the left-right group action. (See e.g. \cite{Schomerus:2005bf} for more explanation, or  \cite{Troost:2011fd,Mitev:2011zza}
for  illustrative diagrams.) In \cite{Schomerus:2005bf} it was further argued that the mini-superspace analysis, and properties
of the correlators suggest that the full conformal field theory spectrum follows the same pattern, with the representations
of the $gl(1|1)$ algebra replaced by primary current algebra representations. The state space $H$ of the Wess-Zumino-Witten model was argued to be
\begin{eqnarray}
H &=& \int_{e \neq 0 \,  \text{mod} \, 1} \, de \, dn \hat{T}_{e,n}^L \otimes \hat{T}^R_{-e,-n} \oplus \sum_m \int \, dq \hat{J}_q^m \, ,
\end{eqnarray}
where $\hat{T}_{e,n}$ are affine generalizations of the typical representations, and $\hat{J}_q^m$ are affine generalizations
of the entangled left-right projective representations. These spaces are described in detail in \cite{Schomerus:2005bf}. (See also
\cite{Creutzig:2011np} for further analysis of the current algebra representation theory.)

{From} the mini-superspace analysis various crucial features are manifest \cite{Schomerus:2005bf}. Firstly, that the scaling operator $L_0$, which on a given
current algebra primary acts as the quadratic Casimir, is not diagonalizable. In particular, it sends top states in projective representations to 
bottom states. Secondly, the chiral spectrum consists of current algebra generalizations of typical and projective representations.

Thus, it is natural to study the model in terms of these current algebra typical and projective representations.
 The typical primaries
 have conformal weight equal to the quadratic Casimir determined by the energy-momentum tensor $T$.
The weight of the $\hat{T}_{e,n}$ primary is
\begin{eqnarray}
h_{e,n} &=&  en + \frac{1}{2} e^2 \, .
\end{eqnarray}
The conformal weight of projective primaries is zero. However, the action of the quadratic Casimir (and of the scaling operator $L_0$) maps the top projective
current algebra primary onto the bottom projective current algebra primary in the same representation.

\subsection{A Free Field Representation}
The $c=-2$ symplectic fermions are akin to a 
$bc$ ghost system with two extra conjugate zero modes  \cite{Flohr:2003tc}. These extra zero modes are crucial to
realize the Jordan block structure of the logarithmic conformal field theory. Thus, in choosing a free field representation
of our theory, we opt for the bosons plus symplectic fermion construction put forward in \cite{Creutzig:2008an}. We remind the reader
of the main formulas.
\subsubsection{The Action and Currents}
The action for the free bosons $Y,Z$ and the symplectic fermions $\chi^{1,2}$ on a Riemann surface $\Sigma$ is
\begin{eqnarray}
S &=& \frac{1}{4 \pi} \int_{\Sigma} d^2 z ( - \partial Z \bar{\partial} Y -
\partial Y \bar{\partial} Z) + \epsilon_{ab} \partial \chi^a \bar{\partial} \chi^b) \, ,
\end{eqnarray}
where we have the convention that $\epsilon_{12}=1$. The action leads to the operator product expansion on the plane
\begin{eqnarray}
\chi^a(z,\bar{z}) \chi^b(w,\bar{w}) & \approx & - \epsilon^{ab} \log |z-w|^2
\nonumber \\
Z(z,\bar{z}) Y(w,\bar{w}) & \approx & \log |z-w|^2 
\, ,
\end{eqnarray}
where $\epsilon^{12}=-1$. We thus also have the operator product expansions
\begin{eqnarray}
\partial Y(z) :e^{e Z}: & \approx & \frac{e}{z-w} :e^{e Z}:(w,\bar{w})
\nonumber \\
\partial Z (z) :e^{n Y}: & \approx &  \frac{n}{z-w} :e^{n Y}:(w,\bar{w}) \, .
\end{eqnarray}
We represent the $\widehat{gl(1|1)}$ affine algebra through the currents
\begin{eqnarray}
E(z) &=&  \partial Y^L \nonumber \\
N(z) &=& \partial Z^L \nonumber \\
\Psi^+(z) &=&   -e^{Y^L} \partial \chi^2 \nonumber \\
\Psi^-(z) &=&  e^{-Y^L} \partial \chi^1 \, . \label{leftcurrents}
\end{eqnarray}
We need twist fields $\mu_\lambda$, introduced in \cite{Kausch:2000fu}
which twist the symplectic fermions $\chi^a$ by a phase determined by $\lambda$:
\begin{eqnarray}
\chi^{1,L}(e^{2 \pi i} z) \mu_{\lambda}(0) &=& e^{- 2 \pi i \lambda} \chi^{1,L}(z) \mu_\lambda(0) \nonumber \\
\chi^{1,R}(e^{-2 \pi i} \bar{z}) \mu_{\lambda}(0) &=& e^{- 2 \pi i \lambda} \chi^{1,R}(\bar{z}) \mu_\lambda(0) \nonumber \\
\chi^{2,L}(e^{2 \pi i} z) \mu_{\lambda}(0) &=& e^{ 2 \pi i \lambda} \chi^{2,L}(z) \mu_\lambda(0) \nonumber \\
\chi^{2,R}(e^{-2 \pi i} \bar{z}) \mu_{\lambda}(0) &=& e^{ 2 \pi i \lambda} \chi^{2,R}(\bar{z}) \mu_\lambda(0) \, .
\end{eqnarray}
We have introduced the chiral parts of the symplectic fermions
\begin{eqnarray}
\chi^a(z,\bar{z}) &=& \chi^{a,L} (z) + \chi^{a,R} (\bar{z}) \, .
\end{eqnarray}
The following operator product expansions were derived in  \cite{LeClair:2007aj,Creutzig:2008an}:
\begin{eqnarray}
\partial \chi^1(z) \mu_\lambda^L(0) & \approx & \frac{1}{z^\lambda} \mu^L_{\lambda-1}(0)
\nonumber \\
\partial \chi^2  \mu_\lambda^L & \approx & \frac{\lambda}{z^{1-\lambda}} \mu^L_{\lambda+1}
\nonumber \\
\bar{\partial} \chi^1 \mu_\lambda^R & \approx & \frac{\lambda}{\bar{z}^{1-\lambda}} \mu^R_{\lambda+1}
\nonumber \\
\bar{\partial} \chi^2  \mu_\lambda^R & \approx & -\frac{1}{\bar{z}^\lambda} \mu^R_{\lambda-1} \, ,
\end{eqnarray}
where we have also split the twist field $\mu$ into holomorphic and anti-holomorphic parts $\mu^{L,R}$.
\subsubsection{Free Field Representations}
We have reviewed the action, the currents and the twist fields. We now recall the construction of the current algebra primary vertex operators.
Firstly, we exhibit the operators associated
to the representations $T_{e,n}$  \cite{Creutzig:2008an}.
We concentrate on the left chirality only, and then find the two-dimensional representation\footnote{Truly, these are infinite dimensional affine
representations $\hat{T}_{e,n}$. We hope the reader will allow  us to leave a multitude of hats on the hat rack.}
\begin{eqnarray}
T_{e,n}^L &=& \left( \begin{array}{cc}
e^{e Z^L + (n-\frac{1}{2}) Y^L} \mu^L_{-e}  &\qquad
 e^{e Z^L + (n+\frac{1}{2})Y^L} \mu^L_{-e+1} 
\end{array} \right) = \left( \, \downarrow \quad \uparrow \, \right) \, . \label{lefttypical}
\end{eqnarray}
The operator product expansions of the currents (\ref{leftcurrents}) with the vertex operator (\ref{lefttypical})
are of the form (\ref{currentprimary}) with the typical representation matrices inserted.
For the projective representation for the left action, we propose (see also \cite{LeClair:2007aj})
\begin{eqnarray}
P^L_{n} &=&  \left( \begin{array}{cc}
 e^{ n Y^L} :\chi^{1,L} \chi^{2,L}: &
- e^{ (n-1)Y^L}\chi^{1,L} \\
 e^{ (n+1) Y^L} \chi^{2,L} & 
- e^{ n Y^L}
\end{array} \right)=
 \left( \begin{array}{cc}
 \mbox{top} &
\mbox{right} \\
\mbox{left} & 
\mbox{bottom}
\end{array} \right)
 \, . \label{leftprojective}
\end{eqnarray}
To obtain the spectrum, we need to assemble these fields into mutually local left-right vertex operators.
For the typical representation, it is sufficient to tensor
\begin{eqnarray}
& & T^L_{e,n} \otimes T^R_{-e,-n} = \nonumber \\
& &   \left( \begin{array}{cc}
e^{e Z + (n-\frac{1}{2}) Y} \mu_{-e} & e^{e Z + (n+\frac{1}{2}) Y^L + (n-\frac{1}{2})Y^R} \mu_{-e+1}^L \mu_{-e}^R
 \\
 e^{e Z + (n-\frac{1}{2}) Y^L + (n+\frac{1}{2})Y^R} \mu_{-e}^L \mu_{-e+1}^R & e^{e Z + (n+\frac{1}{2}) Y} \mu_{-e+1}
\end{array} \right) \, ,
 \nonumber
\end{eqnarray}
and these vertex operators are mutually local for all $e \neq 0$ and all $n$.
 The top component of the left projective representation (\ref{leftprojective})
 is the left part of the vertex
operator $e^{n Y} : \chi^1 \chi^2:$ which is mutually local with the other vertex operators. It is also the top component of a right projective representation.
 In general,  the mutually local primary projective vertex operators
will decompose into the infinite dimensional indecomposable representations $\hat{J}^m_q$ that make up the spectrum at $e=0$.

We use the free field representation of the primary current algebra vertex operators intermittently throughout the rest of the paper.

\section{Logarithmic Aspects of the $gl(1|1)$ WZW Model}
\label{log}
In this section, we discuss consequences of the logarithmic nature of the Wess-Zumino-Witten model
on $gl(1|1)$. We combine the information provided by the representation theoretic content of the model reviewed in section \ref{gl11wzw},
and the global logarithmic conformal Ward identities recalled in section \ref{symmetries}. We treat correlators in increasing order of the number
of insertions.
\subsection{One Point Functions}
The one point functions of conformal primaries are constants by translational invariance. In non-logarithmic
conformal field theories, the global conformal Ward identities imply that all one point functions are zero,
except for operators with conformal dimension zero. In unitary theories, this implies that only the identity
operator has a non-zero one point function in a $SL(2,\mathbb{C})$ invariant vacuum. In the logarithmic
conformal field theory at hand, the standard wisdom is modified. The identity operator
sits inside a Jordan block of rank two. From the conformal scaling Ward identity for the logarithmic partners
$\Phi_0=1$ and $\Psi_0$
in the zero conformal dimension Jordan block
it follows that
\begin{eqnarray}
\hat{\delta}_0 \langle \Psi_0 \rangle =  \langle 1 \rangle &=& 0 \, ,
\end{eqnarray}
where $1$ indicates the unit operator corresponding to the $SL(2,\mathbb{C})$
invariant vacuum. 
The fact that the unit operator has zero expectation value is also a consquence of the existence of two fermionic
zero modes in the $gl(1|1)$ path integral. Thirdly, it follows because the unit operator lies at the bottom
of a projective representation, and therefore is also the result of the quadratic Casimir acting on the corresponding top component.
Applying  the fact that the quadratic Casimir acting on a correlation function is zero, we again derive that the unit operator has
vanishing correlation function. These derivations are all closely related.

The logarithmic partner $\Psi_0$ of the identity operator, on the other hand, can have a non-zero expectation value. The corresponding state contains
two fermionic zero modes conjugate to the global fermionic symmetry generators. These
fermionic modes render the resulting correlation functions non-zero.
From the fact that the identity operator has zero $N$ and $E$ eigenvalues, and the structure of projective representations
under the action of the fermionic generators of $gl(1|1)$, we conclude that it is the top state
in the projective representation $P_0$ that has a non-zero one point function. Indeed, it is the logarithmic
partner $\Psi_0$ of the identity operator $1$. Note though that only the identity operator is a true group invariant inside $P_0$.
By invariance under the bosonic subalgebra of $gl(1|1)$, and zero mode counting, $\langle \Psi_0 \rangle$ 
is also the only non-zero one point function. We can alternatively
summarize these statements in the equations
\begin{eqnarray}
\langle T_{e,n} \rangle &=& 0 \, ,
\nonumber \\
\langle (P_n)_i \rangle &=& \delta_{n} \delta_{i,top} \, ,
\end{eqnarray}
which summarize that the operators in typical representations have zero expectation values, while the top component in the momentum zero
projective representation only can have a non-zero one point function.

These statements are in accord with the free field representations of the current algebra primaries.
In detail, in the symplectic fermion language, the operator $:\chi^1 \chi^2:$ will correspond to
the operator $\omega=:\chi^1 \chi^2:$ of \cite{Kausch:2000fu}, and  is the only operator with a non-zero vacuum expectation value.

\subsection{Two Point Functions}
Typical representation primaries are ordinary (i.e. non-logarithmic) Virasoro primaries.
To have a non-zero two point function amongst two typicals, we need the tensor product
representation to land in the identity Jordan
block, i.e. to combine into a projective cover $P_0$. The top component of the identity block then leads to a non-zero two point 
function. Thus, typical representation primary vertex operators can have
a non-zero two point function when they have total momentum equal to zero, and indeed
form a projective representation. Typical vertex operators are thus pre-logarithmic.
The tensor product of a typical and a projective representation  is typical, and therefore there can be no
non-zero two point function between a typical and a projective representation. This also follows from bosonic
group invariance. A projective representation (with regard to the left action)
contains three ordinary primaries, and one logarithmic partner. The two point function of the logarithmic partners
with themselves is logarithmic, while other combinations that saturate the fermionic zero modes will have
two point functions of standard form. We provide further details in section \ref{correlatorsKZ}.

\subsection{Three Point Functions}
In this section we concentrate on determining the consequences of the logarithmic Ward identities for
 the correlation functions of one projective and two typical current algebra primaries.\footnote{The typical representations can be replaced by
any ordinary Virasoro primary for the reasoning that follows.} See \cite{Flohr:2001tj,Flohr:2001zs,Flohr:2005qq} for the original analysis.
Let's suppose we concentrate on a projective representation primary $\Psi_{h_1}$ that is mapped to
$\Phi_{h_1}$ by the scaling operator $L_0$.\footnote{The projective representation will necessarily have conformal dimension zero. We postpone the use
of this knowledge to illustrate the generality of the reasoning.}
We thus want to study the global conformal Ward identity in the case in which we have a first insertion with a logarithmic partner
\begin{eqnarray}
\hat{\delta}_{h_1} \Psi_{h_1}(z_1) &=& c_1 \Phi_{h_1} (z_1) 
\, ,
\end{eqnarray}
where we allowed for a general normalization constant $c_1$.
We first define a partner three point function of the ordinary Virasoro primaries
\begin{eqnarray}
C^{orig}_{3,0}(z_{1,2,3}) &=& \langle \Phi_{h_1}(z_1) \Phi_{h_2}(z_2)\Phi_{h_3}(z_3) \rangle
\end{eqnarray}
as well as the three point function under study
\begin{eqnarray}
C_{3,1}^{orig}(z_{1,2,3}) &=& \langle \Psi_{h_1}(z_1) \Phi_{h_2}(z_2)\Phi_{h_3}(z_3) \rangle \, .
\end{eqnarray}
The only crucial property of the fields $\Phi_{h_i}$ is that they are ordinary Virasoro primaries.
The lower indices refer to the number of insertions, and the number of logarithmic insertions.
As we do for the standard global conformal Ward identities, we eliminate the conformal dimension $h_i$ dependence
by taking out a pre-factor
\begin{eqnarray}
C_{3,1}^{orig}(z_{1,2,3}) &=& \prod_{i<j} (z_i-z_j)^{\mu_{ij}} C_{3,1}(z_{kl}) \, ,
\end{eqnarray}
where we set
\begin{eqnarray}
\mu_{12} = h_3 -h_1-h_2 \, , \qquad
\mu_{13} = h_2 -h_1-h_3\, , \qquad
\mu_{23} = h_1 - h_2 -h_3 \, .
\end{eqnarray}
The logarithmic global conformal Ward identities then read 
\begin{eqnarray}
\sum_i \partial_i C_{3,1} &=& 0
\nonumber \\
\sum_i z_i \partial_i  C_{3,1} &=& -c_1 C_{3,0}
\nonumber \\
\sum_i z_i^2 \partial_i C_{3,1} &=& -2 z_1 c_1 C_{3,0} \, .
\end{eqnarray}
We split the solution to the logarithmic Ward identities into a homogeneous and a
particular solution. The homogeneous solution, as is familiar from standard three point functions,
is a constant. We thus obtain the  sum
\begin{eqnarray}
C_{3,1} &=& c_{3,1} + C_{3,1}^{part}(z_{kl}) \, ,
\end{eqnarray}
while for the particular solution we adopt the ansatz
\begin{eqnarray}
C_{3,1}^{part} &=& c_{3,0} \sum_{i<j} c_{ij} \log z_{ij} \, ,
\end{eqnarray}
where we used the fact that the rescaled ordinary three point function $C_{3,0}=c_{3,0}$ is a constant.
To find the particular solution then, we have to solve the equations
\begin{eqnarray}
\sum_{i<j} c_{ij} = -c_1 \, ,
\qquad 
 \sum_{j} c_{1j} = - 2 c_1 \, ,
\qquad
c_{12}+c_{23} = 0 = c_{13}+c_{23} \, ,
\end{eqnarray}
which is easily done:
\begin{eqnarray}
-c_{23} = c_{12}=c_{13} = -c_1
\, .
\end{eqnarray}
We have thus derived the chiral three point logarithmic correlator
\begin{eqnarray}
C_{3,1}^{orig} &=& \prod_{i<j} z_{ij}^{\mu_{ij}} (c_{3,1} + c_1 c_{3,0} \log \frac{z_{23}}{z_{12} z_{13}} ) \, ,
\label{solWard31}
\end{eqnarray}
in terms of the constants $c_{3,1},c_{3,0},c_1$ and the conformal dimensions of the operators. This is the form
that the three point function of one top component in a projective representation and two (e.g. typical) Virasoro 
primaries will take.

\subsection{Four Point Functions}
We turn to determining  the form of four point functions. We remind the reader that the standard
global $SL(2,\mathbb{C})$ conformal Ward identities   are solved
by the ansatz:
\begin{eqnarray}
\langle \Phi_{h_1}(z_1) \dots \Phi_{h_n}(z_n) \rangle
&=& \prod_{i < j} (z_i-z_j)^{\mu_{ij}} C_n(z^{kl}_{mp})
\nonumber \\
\sum_{j \neq i} \mu_{ij} &=& -2 h_i \, ,
\end{eqnarray}
where we introduced the function $C_n$ of all cross ratios
\begin{eqnarray}
z_{mn}^{kl} &=& \frac{(z_m-z_p)(z_k-z_l)}{(z_m-z_l)(z_k-z_p)} \, . 
\end{eqnarray}
In the case of four point functions, there is a single independent cross ratio. 
The solution for the powers $\mu_{ij}$ appearing in the prefactor are  not unique.
We can for instance pick the solution:
\begin{eqnarray}
\mu_{12} &=& \mu_{13} = 0 \, , \nonumber \\
\mu_{14} &=& -2 h_1 \nonumber \\
\mu_{23} &=& h_4-h_1-h_2-h_3 \nonumber \\
\mu_{24} &=& h_1+h_3-h_2-h_4  \nonumber \\
\mu_{34} &=& h_1+h_2 - h_3 - h_4 \, .
\end{eqnarray}
We denote the cross ratio by $x$ 
\begin{eqnarray}
x &=& \frac{(z_1-z_2)(z_3-z_4)}{(z_1-z_4)(z_3-z_2)} 
\end{eqnarray}
We can then write the solution to the global conformal Ward identity as
 \begin{eqnarray}
 \langle \Phi_{h_1} \Phi_{h_2} \Phi_{h_3} \Phi_{h_4} \rangle &=& 
\prod_{i<j} z_{ij}^{\mu_{ij}} \, 
 C_{4,0} (x) \,. \label{standardfourpoint}
 \end{eqnarray}
The analysis applies when all four insertions are ordinary primaries of the Virasoro algebra.

We now solve an example of generalized global Ward identities, applicable to a four point function with one logarithmic
insertion and three ordinary
 primaries. We denote the four point function by
\begin{eqnarray}
C^{orig}_{4,1}(z_1,z_2,z_3,z_4) &=&  \langle \Psi_{h_1}(z_1) \Phi_{h_2}(z_2)\Phi_{h_3}(z_3)\Phi_{h_4}(z_4) \rangle \, ,
\end{eqnarray}
and it satisfies the equations
\begin{eqnarray}
\sum_i \partial_i C_{4,1}^{orig} &=& 0
\nonumber \\
(\sum_i z_i \partial_i + (h_i + \hat{\delta}_{h_i})) C_{4,1}^{orig} &=& 0
\nonumber \\
(\sum_i z_i^2 \partial_i + 2 z_i (h_i + \hat{\delta}_{h_i})) C_{4,1}^{orig} &=& 0 \, .
\end{eqnarray}
We again denote and normalize the logarithmic
partner as
\begin{eqnarray}
\hat{\delta}_{h_1} \Psi_{h_1}(z_1) &=& c_1 \Phi_{h_1} (z_1)
\, .
\end{eqnarray}
The equations to solve become
\begin{eqnarray}
\sum_i \partial_i C_{4,1}^{orig} &=& 0
\nonumber \\
(\sum_i z_i \partial_i + h_i ) C_{4,1}^{orig} &=& -c_1 C_{4,0}^{orig}
\nonumber \\
(\sum_i z_i^2 \partial_i + 2 h_i z_i) C_{4,1}^{orig} &=& -2 c_1 z_1 C_{4,0}^{orig}  \, .
\end{eqnarray}
Since the correlator $C_{4,0}^{orig}$ contains only Virasoro primaries, we know that $C_{4,0}^{orig}$ is of the form of the standard four point functions
\begin{eqnarray}
C_{4,0}^{orig} &=& \prod_{i<j} z_{ij}^{\mu_{ij}} 
C_{4,0}(x) \,.
\end{eqnarray}
As a first simplification of the logarithmic four point function, we take out a factor 
\begin{eqnarray}
C_{4,1}^{orig} (z_i) &=& \prod_{i<j} z_{ij}^{\mu_{ij}} 
C_{4,1}(z_{ij}) \, ,
\end{eqnarray}
as well, with $C_{4,1}$ a function of the differences $z_{ij}$ of the positions. In this way, we kill the homogeneous term on the left
hand side, and can cancel the common factors on both sides of the equation to find
\begin{eqnarray}
\sum_i z_i \partial_i  C_{4,1}(z_{jk}) &=& -c_1 C_{4,0}(x)
\nonumber \\
\sum_i z_i^2 \partial_i  C_{4,1}(z_{jk}) &=& -2 c_1 z_1 C_{4,0}(x)  \, .
\end{eqnarray}
To any solution for $C_{4,1}$, we can add a solution to the homogeneous equation, which is a function of the cross ratio.
We thus write
\begin{eqnarray}
C_{4,1} (z_{ij}) &=& C_{4,1}^{hom}(x) + C_{4,1}^{part} (z_{ij}) \, . \label{solWard41}
\end{eqnarray}
We use as an ansatz for the particular solution
\begin{eqnarray}
C_{4,1}^{part} &=& \sum_{i<j} c_{ij} \log z_{ij} C_{4,0}(x)
\, ,
\end{eqnarray}
which leads to the equations
\begin{eqnarray}
\sum_{i<j} c_{ij} = -c_1 \, ,
\qquad
\sum_{j} c_{1j} = -2 c_1 \, ,
\qquad
c_{12}+c_{23}+c_{24} = 0 \, ,
\nonumber \\
c_{13}+c_{23}+c_{34} = 0 \, ,
\qquad
c_{14}+c_{24}+c_{34} =0 \, .
\end{eqnarray}
The generic solution is
\begin{eqnarray}
c_{14} &=& -2 c_1 -c_{12} - c_{13} \nonumber \\
c_{23} &=& -c_1 - c_{12} - c_{13} \nonumber \\
c_{24} &=&  c_1 + c_{13} \nonumber \\
c_{34} &=&  c_1 + c_{12} \, .
\end{eqnarray}
A particular solution is
\begin{eqnarray}
C_{4,1}^{part}(z_i) &=&  c_1 \log \frac{z_{24} z_{34} }{z_{14}^2 z_{23}} C_{4,0}(x) \, . \label{logfourpointpart}
\end{eqnarray}
We note that this particular solution goes to zero for $z_4 \rightarrow \infty$. Moreover, the $z_1$ derivative will also go to zero
when $z_4$ goes to infinity. Therefore, one practical conclusion we draw from our analysis is 
that there is a solution to the logarithmic conformal Ward identities that has the property that
for the choice of points $(z_1,z_2,z_3,z_4)=(x,0,1,\infty)$ the non-homogeneous logarithmic term will not contribute to the Knizhnik-Zamolodchikov equation 
for the $z_1$ derivative of the four point function. We can, so to speak, throw away the ladder after we have climbed it.

\section{Correlation Functions}
\label{correlatorsKZ} 
In the previous section, we analyzed  constraints on the correlation functions that arise from combining the logarithmic
rank two nature of the $gl(1|1)$ Wess-Zumino-Witten model with the global conformal Ward identities. In this section, we study
 the further constraints on the correlation functions implied by the Knizhnik-Zamolodchikov equations. We start out with
two- and three point functions to illustrate the known power of the method, and to ponder the peculiarities of the logarithmic
model at hand. In section \ref{four}, we treat four point functions.

\subsection{Two Point Functions}
For the two point functions, we distinguish two cases. We analyze the two point function of primary operators in typical 
representations, and the two point function of primary operators in projective representations.

\subsubsection{Two typicals}
As remarked previously, two typical representations need to combine into a projective representation in order to find
a non-zero two point function. There will be a single such projective representation, and since the non-top states have zero 
one point function, we find that typical representations
have a single $N,E$ invariant two point function that is non-zero.
The Knizhnik-Zamolodchikov equation for this two point function reads
\begin{eqnarray}
(\partial_1 - \frac{Q_{1 \otimes 2}}{z_1-z_2}) \langle T_{e_1,n_1}(z_1) T_{e_2,n_2}(z_2) \rangle &=& 0 \, .
\end{eqnarray}
We denote by $Q_{i \otimes j}$ the operator $Q_{i \otimes j}= Q_{ab} t^a_i \otimes t^b_j$ acting on
the tensor product representation of the $i$th and $j$th factor in the correlation function.
For a non-zero two point function, we demand invariance under the bosonic generators in the tensor
product representation, and therefore that $e_1+e_2 =0$. The fact that only the top state in the projective
tensor product representation has a non-zero one point function combined with invariance
under the $N$-generator implies that also $n_1+n_2=0$.
The only coefficients of the state that survive projection by the tensor product quadratic Casimir with respect to the left action are given
by
\begin{eqnarray}
\delta_{1 \uparrow} \delta_{2 \downarrow}+\delta_{1 \downarrow} \delta_{2 \uparrow} \, ,
\end{eqnarray}
and therefore the two point function is proportional to this tensor structure, on the left hand side.
Finally, let us calculate its value.
We can use translation invariance to write the Knizhnik-Zamolodchikov equation as
\begin{eqnarray}
(\partial_z - \frac{Q_{1 \otimes 2}}{z}) \langle T_{e_1,n_1} (z) T_{e_2,n_2}(0) \rangle &=& 0 \, .
\end{eqnarray}
Since there is a single top channel in the tensor product, the action of $Q_{1 \otimes 2}$ necessarily
leads to a result proportional to the single top channel up to states that give rise to zero
two point function. The proportionality constant is such that
\begin{eqnarray}
(\partial_z + \frac{h_{e_1,n_1}+h_{e_2,n_2}}{z}) \langle T_{e_1,n_1} (z) T_{e_2,n_2}(0) \rangle &=& 0 \, .
\end{eqnarray}
We recuperate the derivation of the two point function from the global scaling conformal Ward identity,
when it is non-zero. 
Thus, the two point function takes the value
\begin{eqnarray}
 \langle T_{e_1,n_1} (z) T_{e_2,n_2}(0) \rangle &=& c \, \delta_{e_1+e_2} \delta_{n_1+n_2} (\delta_{1 \uparrow} \delta_{2 \downarrow}+\delta_{1 \downarrow} \delta_{2 \uparrow})
z^{-h_{e_1,n_2}-h_{e_2,n_2}}
\, ,
\end{eqnarray}
in as far as the left-moving sector is concerned. We find the full correlator from an identical analysis on the right hand side, and combining them
 consistently with locality.

We can reproduce the resulting correlator using the free field representation and the  symplectic fermion correlator \cite{Kausch:2000fu}
\begin{eqnarray}
\langle \mu_{-e}(z_1,\bar{z}_1) \mu_{1+e}(z_2,\bar{z}_2) \rangle &=& -|z_{12}|^{-2 e(e+1)} \, ,
\end{eqnarray}
as well as the correlator
\begin{eqnarray}
\langle \mu^L_{-e} \mu^R_{-e+1}  (z_1,\bar{z}_1) \mu^L_{1+e}  \mu^R_{e} (z_2,\bar{z}_2) \rangle &=& -z_{12}^{-e(e+1)} \bar{z}_{12}^{-e(e-1)}  \, .
\end{eqnarray}

\subsubsection{Two projectives}
For two projective representations, we again have
\begin{eqnarray}
(\partial_z - \frac{Q_{1 \otimes 2}}{z}) \langle P_{n_1}(z) P_{n_2}(0) \rangle &=& 0 \, .
\end{eqnarray}
There are four top states that can have a non-zero one point function in the tensor product of two projective $P_n$ representations.
For each of the conditions $\sum n_i \pm 1 =0$, there is one combination of states with a non-zero two point function. 
These are mapped to (a state equivalent to) zero by the $Q_{1 \otimes 2}$ 
operator and therefore have a constant two point function. 
We  have two channels when the momenta sum to zero,  $\sum n_i=0$. We refer to appendix \ref{channels} for an analysis of these representation
theoretic facts in all cases of relevance to this paper. Moreover, in appendix \ref{topandinvariants}, we catalogue the
corresponding bottom component invariants.\footnote{These bottom invariants are used as a stand-in for the corresponding top
components. They are handy in that they already incorporate the fact that all non-top components have zero correlation function.
They capture the image of the $C_2^{\otimes}$ map.} We denote these invariants $I_{a_1,a_2;a_3}^b$ where $a_1$ is the number of
typical insertions, $a_2$ the number of projective covers, $a_3$ the negative of the sum of the $n_i$ quantum numbers,
and $b$ a label running over all invariants in the channel.
In the case at hand, the $Q_{1 \otimes 2}$ operator maps the invariant $I_{0,2;0}^2$ into the structure $I_{0,2;0}^1$ times
$2$. Furthermore, we introduce 
the notation $\tilde{I}_{a_1,a_2;a_3}^b$  for the matrix of coefficients of the insertions in the correlator expressed in terms
of the invariants $I_{a_1,a_2;a_3}^b$, after acting by the quadratic Casimir $C_2^\otimes$.

We can then parameterize the two-point function as
\begin{eqnarray}
\langle P_1(z) P_2(0) \rangle &=& G_a^{0,2;0}(z) \tilde{I}_{0,2;0}^a \, ,
\end{eqnarray}
 We  find the coupled system of Knizhnik-Zamolodchikov differential equations
\begin{eqnarray}
\partial_z  G_{1}^{0,2;0} &=& 0 \nonumber \\
\partial_z G_2^{0,2;0} &=& 
2 \frac{G_{1}^{0,2;0}}{z} \, .
\end{eqnarray}
These equations have the solutions
\begin{eqnarray}
G_1^{0,2;0}(z) &=& g_1^{0,2;0}
\nonumber \\
G_{2}^{0,2;0} (z) &=& g_{2}^{0,2;0} + 
2 g_1^{0,2;0}  \log z \, .
\end{eqnarray}
A  short calculation shows that the two top states $\tilde{I}^a_{0,2;0}$ 
indeed form a Jordan block of rank two with respect to the diagonal $L_0$ operator, generating the standard structure
of the two point functions for a logarithmic conformal field theory.

In the free field formalism, at $n_1+n_2=0$, we have the possibility of combining the two top states in the projective
representations into a symplectic fermion two point function 
$\langle :\chi^1 \chi^2 : (z_1,\bar{z}_1) :\chi^1 \chi^2 : (z_2,\bar{z}_2) \rangle$, or the other states in such a
way as to cancel the two fermionic zero modes and obtain $\chi^1 \chi^2$ combinations. In the first
case, we obtain a two point function proportional to  
$-2 (Z+\log |z_{12}|^2)$, where $Z$ is an arbitrary constant, and in the second case, proportional to one \cite{Kausch:2000fu}.

To familiarize the reader further with the algebraic properties of the model,
let us analyze this simplest of examples in even further detail.
From the free field representations, we find the left correlators
\begin{eqnarray}
\langle tt \rangle_L = -2 (\frac{Z}{2}+ \log z_{12})  \, , 
\nonumber \\
\langle lr \rangle_L = 1
\, , \qquad
\langle rl \rangle_L = -1 \, , 
\nonumber \\
\langle tb \rangle_L = -1
\, , \qquad
\langle bt \rangle_L = -1 \, .
\end{eqnarray}
We can check that for instance the Knizhnik-Zamolodchikov equation
\begin{eqnarray}
\partial_1 \langle tt \rangle_L + \langle lr-rl \rangle_L/z_{12} &=& 0
\end{eqnarray}
is satisfied. We also find that the second Casimir acting on the $\langle tt \rangle_L$ correlator
gives rise to a zero correlator
\begin{eqnarray}
C_2 \langle tt \rangle_L &=& - \langle bt + lr -rl + tb \rangle_L =0 \, ,
\end{eqnarray}
as required.  In analyzing the consequences of the
Knizhnik-Zamolodchikov equation, we mostly concentrate on the
holomorphic, left-moving correlators. It will be manifest when we
switch back to the full non-chiral correlators. We often drop the
subscript $L$ from the chiral correlators for ease of notation.

\subsection{Three Point Functions}
For the three point functions, we concentrate on an example correlator involving one
projective and two typical representations. The corresponding  Knizhnik-Zamolodchikov equation is
\begin{eqnarray}
(\partial_1 - \frac{Q_{1 \otimes 2}}{z_1-z_2}- \frac{Q_{1 \otimes 3}}{z_1-z_3}) \langle P_{n_1} T_{e_2,n_2} T_{e_3,n_3} \rangle &=& 0 \, .
\end{eqnarray}
In the most interesting $\sum n_i =0$ channel, 
we have two invariants (see appendix \ref{topandinvariants}), and we find the equation
\begin{eqnarray}
(\partial_1 - \frac{Q_{12}}{z_1-z_2}- \frac{Q_{13}}{z_1-z_3}) \left( \begin{array}{c} G_1 \\ G_2 \end{array} \right) &=& 0
\end{eqnarray}
for the position dependent correlators $G_{1,2}$ corresponding to the invariants $I^{1}_{2,1,0}$ and $I^2_{2,1,0}$.
The matrices $Q_{12}$ and $Q_{13}$ are recorded in appendix \ref{Qs}.
We find the solution
\begin{eqnarray}
G_2 &=& g_2 (z_1 - z_2)^{e_2 n_1}(z_1 - z_3)^{ e_3 n_1} \, ,
\end{eqnarray}
consistent with the global conformal Ward identity for the non-logarithmic mode in the projective representation.
Analyzing the $z_2$ dependent Knizhnik-Zamolodchikov equation or the global conformal Ward identity would
determine the $z_{23}$ dependence as well.
We also find the correlator
\begin{eqnarray}
G_1 &=& (z_1 - z_2)^{e_2 n_1}(z_1 - z_3)^{e_3 n_1} (g_1 - g_2 e_2  \log(z_1-z_2)(z_1-z_3)) \, , \nonumber 
\end{eqnarray}
again in accord with the global conformal Ward identity (\ref{solWard31}).
\subsection{A Free Field Representation}
Using the free field representation of the $\langle PTT \rangle$  correlator, we can determine a coherent set of constants in the correlation
functions. This calculation was performed in the $bc$ ghost system in \cite{Schomerus:2005bf}.
The  free bosonic $Y,Z$ correlators that occur are
\begin{eqnarray}
\langle \prod e^{e_i^L Z^L + e_i^R Z^R + n_i^L Y^L+ n_i^R Y^R} \rangle
&=& \prod_{i < j} (z_i-z_j)^{e_i^L n_j^L+e_j^L n_i^L} (\bar{z}_i - \bar{z}_j)^{e_i^R n_j^R+e_j^R n_i^R} \, .
\end{eqnarray}
The relevant symplectic fermion correlators are reviewed in appendix \ref{symplecticfermions}.
For the $\langle PTT \rangle$ correlators, there are $4 \times 2 \times 2$ possibilities for left chiral correlators.
We concentrate again on the channel $\sum n_i =0$. We obtain a first free field correlator
\begin{eqnarray}
\langle t \uparrow \downarrow \rangle &=& z_{12}^{n_1 e_1} z_{13}^{n_1 e_3} z_{23}^{e_2(n_3-\frac{1}{2})+e_3(n_2+\frac{1}{2})}
 (z_{23})^{e_2(1-e_2)} (\frac{1}{2} Z_{e_2} + \log \frac{z_{12} z_{13}}{z_{23}}) \, ,
\end{eqnarray}
where the constant $Z_{e_2}$ equals \cite{Kausch:2000fu}
\begin{eqnarray}
Z_{e_2} &=& Z + 2 \psi(1) - \psi(e_2) - \psi(1-e_2) \, ,
\end{eqnarray}
and $\psi(z)$ is the digamma function.
The  $\langle t \downarrow \uparrow \rangle$ correlator is found by exchanging the second and third entry.
Moreover, we find the left correlators 
\begin{eqnarray}
\langle b \uparrow \downarrow \rangle = 1
\, , \qquad
& &
\langle b  \downarrow \uparrow \rangle = 1
\\
\langle l \downarrow \downarrow  \rangle = -1
\, , \qquad
& & 
\langle r \uparrow \uparrow  \rangle = \frac{1}{e_2} \, ,
\end{eqnarray}
where we left out the canonical $z_{ij}$ dependence for ease of notation.
A correlation function is zero if it is not top (after tensor product decomposition),
which gives rise to the constraints
\begin{eqnarray}
\langle l \downarrow \downarrow \rangle + \langle b \downarrow \uparrow \rangle  = 0 \, , \qquad
& & 
\langle b \uparrow \downarrow \rangle - \langle b \downarrow \uparrow  \rangle = 0 \, ,
\nonumber
 \\
e_2 \langle r \uparrow \uparrow \rangle  - \langle b \downarrow \uparrow  \rangle = 0\, , \qquad
& & 
e_2 (\langle t \uparrow \downarrow \rangle  - \langle t \downarrow \uparrow \rangle ) = \langle b \downarrow \uparrow  \rangle \, .
\end{eqnarray}
To check the last equation, we use the shift identity of the digamma function $\psi$
\begin{eqnarray}
\psi(z+1) &=& \psi(z) + \frac{1}{z} \, .
\end{eqnarray}
We can also check the Knizhnik-Zamolodchikov equations as applied to individual correlators more directly.
An representative example equation is 
\begin{eqnarray}
\partial_1 \langle t \uparrow \downarrow \rangle + \frac{1}{z_{12}} \langle l \downarrow \downarrow \rangle
- \frac{e_2 n_1}{z_{12}} \langle t \uparrow \downarrow \rangle
- \frac{e_2 \langle r \uparrow \uparrow \rangle}{z_{13}} - \frac{e_3 n_1}{z_{13}} \langle t \uparrow \downarrow \rangle &=& 0 \, , \nonumber
\end{eqnarray}
which is indeed satisfied.

 We conclude with the remark that the Knizhnik-Zamolodchikov equations that we encountered
are essentially identical to the differential equations on symplectic fermion correlators recorded in appendix B of \cite{Kausch:2000fu}.
The generic proof of this fact will follow from gauging the bosonic directions in the $gl(1|1)$ Wess-Zumino-Witten model. This structural
map is valid for the four point function of typical representations to be analyzed next, and will also hold for the four point function involving
one projective representation. The latter case therefore adds to our knowledge of symplectic fermion correlators.

\section{The Four Point Functions}
\label{four}
We consider the four point functions in which there are zero or one projective representations amongst
the initial current algebra primaries, and four or three typicals. In the first case, we closely mimic the
original derivation of the four point function \cite{Rozansky:1992rx}. In section \ref{local} we will perform a more general
check of crossing symmetry.  The second case will have features reminiscent of the calculation of correlators
in other models with supergroup symmetries \cite{Maassarani:1996jn}.

\subsection{Four Typicals} 
We first solve for the four typical representations
four point function.  These typical representations are ordinary Virasoro primaries
and as such their correlation functions satisfy the standard (i.e. non-logarithmic)
global conformal Ward identities. Thus, their four point functions are of the form
(\ref{standardfourpoint}).
We consider the Knizhnik-Zamolodchikov equation with respect to the first
insertion at $z_1$. We wish to write the equation as a differential equation in the cross ratio $x$.
The derivative with respect to $z_1$ gives terms arising from
the prefactor, and we  use the chain rule to find
\begin{eqnarray}
\partial_1 C_4(x) &=& (\frac{x}{z_{12}}-\frac{x}{z_{14}}) \partial_x C_4 \, .
\end{eqnarray}
We can then put $z_1=x,z_2=0,z_3=1,z_4=\infty$ to obtain the differential equation
\begin{eqnarray}
( \partial_x  
- \frac{Q_{1 \otimes 2}}{x} - \frac{Q_{1 \otimes 3}}{x-1}) C_4(x) &=& 0
\, .
\end{eqnarray}
The representation theoretic structure of the correlators is again exhibited in appendix \ref{reps}.
We must satisfy the $E$-invariance constraint
 $\sum_{i=1}^4 e_i=0$, and find 
projective representations $P_0$
if we have $\sum n_i = \pm 1$ or $\sum n_i=0$.
These cases give rise to either a single or to two channels.
\subsubsection{Four Typicals  with $\sum n_i+1=0$} 
We concentrate first on a case with 
a single channel, namely $\sum n_i +1=0$.
We wish to evaluate the $Q_{i \otimes j}$ operator on the top state in this channel.
Since there is a single channel, there is only a proportionality constant $q_{1j}$ to compute, which for $Q_{1j}$ is  equal to
\begin{eqnarray}
q_{1j} &=& h_{e_{1}+e_j,n_{1}+n_j+1/2} - h_{e_1,n_1} - h_{e_j,n_j} \, ,
\end{eqnarray}
where the $h_{e,n}$ are conformal dimensions of primaries.
In the tensor product $T_{e_1,n_1} \otimes T_{e_j,n_j}$, we project onto the relevant channel (with highest $N$ eigenvalue --
see appendix \ref{Qs} for details).
We then find the equation \begin{eqnarray}
( \partial_x - \frac{q_{12}}{x} 
-\frac{q_{13}}{x-1}  
) C_{4,0;0}(x) &=& 0
\, .
\end{eqnarray}
We solve the first order differential equation to find the chiral correlator
\begin{eqnarray}
C_{4,0;0} (x) &=& c_{4,0;0} \,  x^{q_{12}} (1-x)^{q_{13}} \, ,
\end{eqnarray}
with $c_{4,0;0}$ an arbitrary integration constant.

\subsubsection{Four Typicals  with $\sum n_i=0$}
In the case where the $N$-momenta sum to zero, 
we have two channels. The action of the operators $Q_{1 i}$ in our basis is
recorded in appendix \ref{Qs}.
The Knizhnik-Zamolodchikov equation becomes a  combination of two equations. We solve one equation for one of the correlators,
substitute it in the other equation, and then solve the resulting hypergeometric equation \cite{Rozansky:1992rx}.
 We can do this for both channels, and then again impose consistency with the original
equations relating the chiral correlators.  We find:
\begin{eqnarray}
G_1 &=& (x-1) (1-x)^{\frac{1}{2} \left(2 e_1 \left(e_3+n_3\right)+2 e_3 n_1+e_1+e_3\right)} x^{e_2 \left(n_1-\frac{1}{2}\right)+e_1 \left(e_2+n_2-\frac{1}{2}\right)} \nonumber 
\\
& & \Big(c_1 
\frac{e_1+e_2+1}{e_2}  x^{e_1+e_2} \, _2F_1\left(e_1+1,e_1+e_2+e_3+1;e_1+e_2+1;x\right) \nonumber
 \\
& & +c_2 \frac{ e_3}{e_1+e_2} \, _2F_1\left(1-e_2,e_3+1;-e_1-e_2+1;x\right) \Big)
\\
G_2 &=& (1-x)^{\frac{1}{2} \left(2 e_1 \left(e_3+n_3\right)+2 e_3 n_1+e_1+e_3\right)} x^{\frac{1}{2} e_2 \left(2 n_1-1\right)+e_1 \left(e_2+n_2-\frac{1}{2}\right)} 
\nonumber \\
& & \Big( c_1 x^{e_1+e_2+1} \,
   _2F_1\left(e_1+1,e_1+e_2+e_3+1;e_1+e_2+2;x\right)
\nonumber \\
& & 
+c_2 \, _2F_1\left(-e_2,e_3;-e_1-e_2;x\right) \Big)  \, . \nonumber 
\end{eqnarray}
 We will later work in terms of the basis where $G_1^1$ is the coefficient of $c_1$ in $G_1$ et cetera.
In section \ref{local} we assemble these chiral building blocks into a left-right local correlation function satisfying crossing
symmetry.

\subsection{One Projective and Three Typicals}
In the second part of this section, we compute the chiral correlators for a four point function containing one projective
and three typical representations.

\subsubsection{One Projective and Three Typicals with $\sum n_i +\frac{1}{2}$=0}
In simplifying the Knizhnik-Zamolodchikov equation, we can pick the solution to the global
conformal Ward identity in the form (\ref{logfourpointpart}) such that when we fix the points $z_i=(x,0,1,\infty)$, 
the potential supplementary logarithmic
contribution vanishes. Alternatively,
we can directly analyze the Knizhnik-Zamolodchikov equation without using the global conformal Ward identity, and only fix the 
points $z_i$ in a later stage. In the following, we choose one or the other technique, depending on convenience.

In any case, with the constraint $\sum n_i +1/2=0$, we find the three channels given in the appendix, equation (\ref{threechannels}),
 and the action of the operators  $Q_{1 \otimes 2}$ and $Q_{1 \otimes 3}$ in the Knizhnik-Zamolodchikov equations are then as
 in  equation (\ref{threechannelsQ}). We note that $L_0$ is not diagonalizable on the invariants. It has non-trivial Jordan structure,
 with equal diagonal entries and mapping the first to the second invariant. 
Fixing the insertion points, we find the three Knizhnik-Zamolodchikov  differential equations
\begin{eqnarray}
-\frac{{e_2 n_1 {G_1}(x)}+ {G_2}(x)}{x}-\frac{e_3 n_1 {G_1}(x)-{e_3
   {G_3}(x)}-e_3/e_2 G_2(x)}{x-1}+{G_1}'(x) &=& 0
\nonumber \\
-\frac{e_2(n_1+1) G_3(x)}{x}-\frac{ (e_2+e_3)/e_2 G_2(x) + {e_3 \left(n_1+1\right) {G_3}(x)}}{x-1}
+{G_3}'(x) &=& 0
\nonumber \\
-\frac{e_2 n_1 {G_2}(x)}{ x}-\frac{e_3 n_1 {G_2}(x)}{ x-1}+{G_2}'(x) = 0 &&  \, .
\end{eqnarray}
We can solve these differential equations consecutively, from simple to hard, and find the solution
\begin{eqnarray}
G_2 (x) &=& (1-x)^{e_3 n_1} x^{e_2 n_1} c_3
\nonumber \\
G_3(x) &=&
(1-x)^{e_3 (n_1+1)} x^{e_2 (n_1+1)} (c_2 - B(x,1-e_2,-e_3) c_3 (e_2+e_3)/e_2)
\nonumber \\
G_1(x) &=&  (1-x)^{e_3 n_1} x^{e_2 n_1} (c_1 + c_3  \log x - c_3 \frac{e_3}{e_2} \log (1-x) 
\nonumber \\
& & 
+ \frac{e_3}{1+e_2}
x^{1 + e_2} c_2 \, {}_2 F_1 (1+e_2,1-e_3,2+e_2;x)\nonumber \\
&& + \frac{e_3(e_2+e_3)}{e_2 (1-e_2)} c_3 x \Gamma(2-e_2) \, {}_p \tilde{F}_q(\{ 1,1-e_2-e_3,1 \} , \{ 2-e_2 , 2 \} ; x)\nonumber \\
&& - c_3 \frac{e_3(e_2+e_3)}{e_2(1-e_2)} \frac{\Gamma(2-e_2)}{\Gamma(1-e_2)\Gamma(1+e_3) \Gamma(1-e_2-e_3)} \nonumber \\
&&
G( \{ \{ 0, e_2+e_3 \} , \{ 1 \} \} , \{ \{ 0,0,e_3 \} , \{ \} \} ;1-x) \, . \label{3KZ}
\end{eqnarray}
We made use of standard notations for the arguments of the regularized generalized hypergeometric function ${}_p \tilde{F}_q$ and the 
Meijer $G$-function which we  denoted $G$. We named the incomplete beta-function $B$.
These solutions contain all information on the chiral correlators.

We found it useful to gain more  insight into how their staircase structure arises
from $gl(1|1)$ representation theory.
There are $4 \times 2^3=32$ $\langle PTTT \rangle$ chiral correlation functions. Let's concentrate on the current primary
states that satisfy the $N$-channel constraint $\sum n_i +\frac{1}{2}=0$ and $N$-invariance. We have ten combinations satisfying these conditions:
\begin{eqnarray}
b \downarrow \uparrow  \uparrow, b \uparrow  \downarrow  \uparrow, b \uparrow   \uparrow \downarrow,
t\downarrow \uparrow  \uparrow, t \uparrow  \downarrow  \uparrow, t \uparrow   \uparrow \downarrow,
l \uparrow \downarrow \downarrow, l \downarrow \uparrow \downarrow,  l\downarrow \downarrow \uparrow,
r \uparrow \uparrow \uparrow  \, .
\end{eqnarray}
We find constraints from the demand that all combinations that map to zero under the quadratic Casimir $C_2$ have
zero correlator. By the channel count, this should leave us with only three non-zero combinations. The
seven constraints are 
\begin{eqnarray}
e_3 \langle  b \downarrow \uparrow \uparrow  \rangle + e_2 \langle b \uparrow \downarrow \uparrow \rangle &=& 0 
\nonumber \\
e_2 \langle b \uparrow \uparrow \downarrow \rangle  - e_4 \langle b \downarrow \uparrow \uparrow \rangle  &=& 0 \label{bottomconstraints} \\
e_2 \langle r \uparrow \uparrow \uparrow \rangle  -\langle  b \downarrow \uparrow \uparrow \rangle &=& 0 \label{rightconstraint} \\
\langle b \downarrow \uparrow \uparrow  \rangle -\langle l \downarrow \uparrow \downarrow \rangle  +\langle  l \downarrow \downarrow \uparrow \rangle &=& 0  \label{leftconstraint1}  \\
e_2\langle  l \uparrow \downarrow \downarrow \rangle  + e_3\langle  l \downarrow \uparrow \downarrow \rangle  + e_4\langle  l \downarrow \downarrow \uparrow \rangle  \label{leftconstraint2} &=& 0 \\
-\langle  t \uparrow \downarrow \uparrow \rangle  +\langle  t \downarrow \uparrow \uparrow  \rangle +\langle  t \uparrow \uparrow \downarrow  \rangle + \frac{1}{e_2}\langle  b \downarrow \uparrow \uparrow \rangle  &=& 0
\label{topbottomconstraint} \\
e_2\langle  t \uparrow \downarrow \uparrow  \rangle + e_3 \langle t \downarrow \uparrow \uparrow \rangle  +\langle  l \downarrow \downarrow \uparrow \rangle &=& 0 
\label{topleftconstraint}
\, .
\end{eqnarray}
To study the staircase structure of the Knizhnik-Zamolodchikov equations, 
we first consider the correlator $\langle b \uparrow \downarrow \uparrow \rangle$
and its permutations in the last three entries. These correlators must satisfy the constraints
(\ref{bottomconstraints}). We first perform a check on our formalism using the free
field representation of the correlator. It consists of a free field $YZ$ correlator,
and a three point function of symplectic fermion twist fields. For the diagonal combination,
the full non-chiral correlator is given by
\begin{eqnarray}
C_{b  \downarrow \uparrow \uparrow}^{ff} &=& \langle e^{n_1 Y} e^{e_2 Z + (n_2-\frac{1}{2}) Y} e^{e_3 Z + (n_3+\frac{1}{2}) Y} e^{e_4 Z + (n_4+\frac{1}{2}) Y} \rangle
\langle \mu_{-e_2} \mu_{-e_3+1} \mu_{-e_4+1} \rangle
\nonumber \\
&=& C_{-e_2,-e_3+1,-e_4+1} |z_{23}^{e_3 (1+e_2)} z_{34}^{e_3 e_4} z_{24}^{(1+e_2) e_4}|^2 
|z_{12}^{n_1 e_2} z_{13}^{n_1 e_3} z_{14}^{n_1 e_4}|^2  \\
& &| z_{23}^{e_2 (n_3+\frac{1}{2})+e_3 (n_2-\frac{1}{2})} z_{24}^{e_2(n_4+\frac{1}{2})+e_4(n_2-\frac{1}{2})} z_{34}^{e_3 (n_4+\frac{1}{2})+e_4 (n_3+\frac{1}{2})}|^2 \, .
\nonumber
\end{eqnarray}
The three point function $C$ determined in \cite{Kausch:2000fu} picks up factors under the interchange of the insertions $2,3,4$.
This becomes manifest when we rewrite
\begin{eqnarray}
C_{-e_2,-e_3+1,-e_4+1} &=& \sqrt{ \Gamma( \frac{e_2,e_3,e_4}{-e_2,-e_3,-e_4}) \frac{e_2}{e_3 e_4} } \, .
\end{eqnarray}
Permuted three point functions differ by factors of $\sqrt{e_3^2/e_2^2}$ or $\sqrt{e_4^2/e_2^2}$.
With the appropriate sign choices (which we make), we then satisfy the constraints (\ref{bottomconstraints}).
The correlator involving the right state $r$ is then fixed through the constraint (\ref{rightconstraint}). All these correlators
sit on the first step of our staircase.

Next, we use the Knizhnik-Zamolodchikov equation to determine the $ \langle l \downarrow \downarrow \uparrow \rangle $ correlator.
The simplest choice of prefactor $\prod_{i<j} z_{ij}^{\mu_{ij}}$ to extract from the differential equation corresponds again to the choice
of $\mu_{ij}$
\begin{eqnarray}
\mu_{12} =0, \quad \mu_{23}=h_4-h_1-h_2-h_3, \quad \mu_{24}=h_1+h_3-h_2-h_4, & & \nonumber \\
\mu_{13}=0, \quad \mu_{14}=- 2 h_1, \quad \mu_{34}=h_1+h_2-h_3-h_4     \, ,     & &
\end{eqnarray}
and we have that $h_1=0$, such that the factors agree with those of an ordinary three point function.
Concretely, before fixing the insertion points, 
we obtain for example the Knizhnik-Zamolodchikov equation
\begin{eqnarray}
\partial_1 \langle l \downarrow \downarrow \uparrow \rangle &=& \frac{1}{z_{12}} ( e_2 \langle b \uparrow \downarrow \uparrow \rangle + e_2(n_1+1) \langle l \downarrow \downarrow \uparrow \rangle)
+ \frac{1}{z_{14}} e_4 (n_1+1) \langle l \downarrow \downarrow \uparrow \rangle
\nonumber \\
& & + \frac{1}{z_{13}} ( e_3 \langle b \downarrow \uparrow \uparrow \rangle + e_3 (n_1+1) \langle l \downarrow \downarrow \uparrow \rangle)
 \, .
\end{eqnarray}
Factoring out the $z_{ij}^{\mu_{ij}}$ dependence, and denoting the chiral $\langle b  \downarrow \uparrow \uparrow \rangle$ correlation
function (up to this factor) by the constant $c_{ \downarrow \uparrow \uparrow}$, we obtain from these and analogous differential
equations the results
\begin{eqnarray}
\langle l  \downarrow \downarrow \uparrow \rangle &=& (1-x)^{e_3 n_1} x^{e_2 n_1} ( (-1)^{1+e_2 n_1} c_{ \downarrow \uparrow \uparrow} + (1-x)^{e_3} x^{e_2} d_1
 \\
& & 
+ (-1)^{e_2 n_1} c_{ \downarrow \uparrow \uparrow} (x-1) \Gamma(-e_2) {}_2 \tilde{F}_1 (1,1-e_2-e_3,1-e_2;x) (e_2+e_3))
\nonumber \\
\langle l \downarrow \uparrow \downarrow \rangle &=& (1-x)^{e_3 (n_1+1)} x^{e_2(n_1+1)} (d_2 + (-1)^{e_2 n_1} c_{ \downarrow \uparrow \uparrow} B(x;-e_2,1-e_3) e_4)
\nonumber \\
\langle l  \uparrow \downarrow \downarrow \rangle &=& x^{e_2 (n_1+1)} (1-x)^{e_3 (n_1+1)} (d_3 + (-1)^{1+e_2 n_1} c_{ \uparrow \downarrow \uparrow} B(x;1-e_2,-e_3) e_4) \, .
\nonumber
\end{eqnarray}
We introduced integration constants $d_i$.
We note the special function identity between the incomplete beta function and the regularized hypergeometric function
\begin{eqnarray}
B(z;a,b) &=& \Gamma(a) (1-z)^b z^a \, {}_2 \tilde{F}_1 (1,a+b;a+1;z)
\end{eqnarray}
which allows to rewrite the first correlator as
\begin{eqnarray}
\langle l  \downarrow \downarrow \uparrow \rangle &=& (1-x)^{e_3 n_1} x^{e_2 n_1} ( (-1)^{1+e_2 n_1} c_{ \downarrow \uparrow \uparrow} +  d_1 (1-x)^{e_3} x^{e_2}
\nonumber \\
& & 
+ (1-x)^{e_3} x^{e_2} (-1)^{e_2 n_1} c_{ \downarrow \uparrow \uparrow} B(x;-e_2,1-e_3) e_4) \nonumber \, .
\end{eqnarray}
These correlators satisfy the equality (\ref{leftconstraint1})
provided $d_1=d_2$. The constraint (\ref{leftconstraint2}) then holds 
if we also impose
$d_3 = d_1$.
To verify the latter result, we use properties of the incomplete beta function under shifts of its arguments
\begin{eqnarray}
B(x,1-e_2,-e_3) &=& \frac{e_2}{e_2+e_3} B(x;-e_2,-e_3) + \frac{1}{e_2+e_3} x^{-e_2} (1-x)^{-e_3}
\nonumber \\
B(x,-e_2,1-e_3) &=& \frac{e_3}{e_2+e_3} B(x;-e_2,-e_3) - \frac{1}{e_2+e_3} x^{-e_2} (1-x)^{-e_3}
\, . \nonumber 
\end{eqnarray}
Alternatively, we can link the left correlators to the invariants we computed previously, and write the correlators in terms
of the functions $G_i$. This links the integration constants $c_2,c_3$ to the constants $d_i,c_{\downarrow \uparrow \uparrow}$
used above. The following two correlators are
sufficient to perform the matching
\begin{eqnarray}
\langle b \downarrow \uparrow \uparrow \rangle &=& - \frac{e_2}{e_3} c_3 x^{e_2 n_1} (1-x)^{e_3 n_1}
\nonumber \\
\langle l \uparrow \downarrow \downarrow \rangle &=& (1-x)^{e_3 (1+n_1)} x^{e_2(1+n_1)}
(-c_2 e_2 - B (x, 1 - e_2, -e_3) c_3 e_4) \, . \nonumber
\end{eqnarray}
We have understood the constraint equations and the properties of our correlators at the second step
of our Knizhnik-Zamolodchikov staircase. We climb the last step in our study of the 
top correlators. An example Knizhnik-Zamolodchikov equation for a top correlator is
\begin{eqnarray}
\partial_1 \langle t\uparrow \uparrow \downarrow \rangle &=& -\langle l \downarrow \uparrow \downarrow \rangle/z_{12} + n_1 e_2\langle t\uparrow \uparrow \downarrow \rangle/z_{12}
+\langle l \uparrow \downarrow \downarrow \rangle/z_{13} + n_1 e_3\langle t\uparrow \uparrow \downarrow \rangle/z_{13} \nonumber \\
& & + e_4\langle r\uparrow \uparrow \uparrow \rangle/z_{14} + n_1 e_4\langle t\uparrow \uparrow \downarrow \rangle/z_{14} \, .
\label{atopKZ}
\end{eqnarray} 
The relation between the top correlators and the coefficients of the (top partners of the)
invariants is
\begin{eqnarray}
 \langle t\uparrow \uparrow \downarrow \rangle &=& \frac{e_2+e_3}{e_3} G_1 + G_3
 \label{atopcorrelator} \\
 \langle t \uparrow \downarrow \uparrow  \rangle&=& G_1 + G_3
\nonumber \\
 \langle t \downarrow \uparrow \uparrow \rangle &=& - \frac{e_2}{e_3} G_1 + \frac{1}{e_3} G_2 \nonumber
\, .
\end{eqnarray}
{From} these equations we can check  the relation
\begin{eqnarray}
\langle t\uparrow \uparrow \downarrow  \rangle+\langle t \downarrow \uparrow \uparrow  \rangle-\langle t \uparrow \downarrow \uparrow   \rangle&=& 
\frac{c_3}{e_3} (1 - x)^{e_3 n_1} x^{e_2 n_1} = -\frac{1}{e_2} \langle b \downarrow \uparrow \uparrow   \rangle
\end{eqnarray}
and the final constraint (\ref{topleftconstraint}) is satisfied. All top correlators involve the generalized hypergeometric
function incorporated in the special function $G_1$. We checked that the combination (\ref{atopcorrelator}) satisfies the
Knizhnik-Zamolodchikov equation (\ref{atopKZ}). We thus reconstructed step-by-step the chiral correlation functions 
summarized in the beginning of this subsection.

\section{Local Correlators}
\label{local}
In this section, we glue together the chiral four point correlators of section \ref{four} to obtain local
four point correlation functions of the conformal field theory. 
\subsection{Four Typicals at $\sum n_i+1=0$}
A typical diagonal four point function is given by the modulus square of the diagonal holomorphic result
\begin{eqnarray}
C_{4,0;1}(x) 
&=& c |x|^{2 q_{12}} |1-x|^{2q_{13}} 
\, .
\end{eqnarray}
We see that $2 \rightarrow 3$ interchange is equivalent to $x \leftrightarrow 1-x$ interchange in the function $C(x)$.
That is a check on crossing symmetry.
\subsection{Four Typicals  at $\sum n_i=0$}
We approach the correlator in two ways. We first determine a local four point function
by analyzing single-valuedness of a proposed left-right correlator near $x=0$ and $x=\infty$. In a second approach,
we check crossing symmetry of the correlator under the interchange of insertions $2$ and $3$.
\subsubsection{Locality}
We can determine the four point function of four typical representations by demanding locality at $x=0$ and $x=\infty$
(and consequently, at $x=1$). To that end, we use the hypergeometric function transformation formulas\footnote{We often denote a ratio of products of Gamma-functions
 by $\frac{\Gamma(a_1)\dots \Gamma(a_n)}{\Gamma(b_1)\dots \Gamma(b_n)}=\Gamma(\frac{a_1,\dots,a_n}{b_1,\dots,b_n})$.}
\begin{eqnarray}
F(a,b,c;z) &=&
\Gamma(\frac{c,b-a}{b,c-a}) (-z)^{-a}
F(a,1-c+a,1-b+a;z^{-1})
\nonumber \\
& & 
+
\Gamma(\frac{c,a-b}{a,c-b}) (-z)^{-b}
F(b,1-c+b,1-a+b;z^{-1}) \, .
\end{eqnarray}
For convenience, we concentrate on the second channel only. We work in the set-up where $e_2 + e_3 \neq 0$.
We can identify combinations of left- and right-moving blocks that are local at $x=0$ (which 
consists of linear combinations of the modulus of $G_2^1$ and the modulus of $G_2^2$), with the
modulus of the combinations local at $x=\infty$.
We find that the combination that is both local at zero and at infinity, equal to
\begin{eqnarray}
G(x,\bar{x}) &=& X_{11} |G_2^1|^2 + X_{22} |G_2^2|^2
\end{eqnarray}
has a ratio of coefficients
\begin{equation}
X_{11}/X_{22} = - 
\Gamma(\frac{1+e_1,1-e_3,-e_1-e_2,1+e_2,1+e_1+e_2+e_3,-e_1-e_2}{-e_1,e_3,2+e_1+e_2,2+e_1+e_2,-e_2,-e_1-e_2-e_3}) \, .
\label{localityconstraint}
\end{equation}

\subsubsection{Crossing Symmetry}
It will be interesting to further study explicitly whether the four point function satisfies crossing symmetry.
 The four point function is given by
\begin{eqnarray}
G(x,\bar{x}) &=& \sum_{i,j=1,2} \tilde{I}_i \tilde{\bar{I}}_j G_{i,j}(x,\bar{x})
\nonumber \\
&=& \sum_{i,j=1,2} \tilde{I}_i \tilde{\bar{I}}_j \sum_{n,m=1,2} X_{nm} G_i^n (x) G_j^{m}(\bar{x}) \, ,
\end{eqnarray}
where $\tilde{I}_i$ denote again the matrix of coefficients of the two (top partners of the) invariants in this channel, and $G_i^n(x)$ the holomorphic
solutions to the Knizhnik-Zamolodchikov equations. We will immediately exploit that
for single-valuedness near $x=0$, we must have that $X_{nm}$ is diagonal in $1,2$.
We thus have
\begin{eqnarray}
G(x,\bar{x})
&=& \sum_{i,j=1,2} \tilde{I}_i \tilde{\bar{I}}_j \sum_{n=1,2} X_{nn} G_i^n (x) G_j^{n}(\bar{x}) \, .
\end{eqnarray}
Crossing symmetry requires that 
\begin{eqnarray}
G(1-x,1-\bar{x};32) &=& G(x,\bar{x};23) \, ,
\end{eqnarray}
where we indicated explicitly that we should exchange all quantities associated to insertions $2$ and $3$ in the
correlation function.
It is important to note now that exchanging insertions $2$ and $3$ has the effect of exchanging the two invariants 
of the global symmetry group that we handily chose in appendix \ref{fourtypicalsinvariants}.
We thus must have that 
\begin{eqnarray}
G_{i,j}(x, \bar{x};23) &=& G_{3-i,3-j}(1-x,1-\bar{x};32) \, . 
\end{eqnarray}
To compute the transformation rule for our holomorphic or anti-holomorphic constituents, 
we use the hypergeometric function transformation formulas
\begin{eqnarray}
F(a,b;c;x) &=& \frac{\Gamma(c) \Gamma(c-a-b)}{\Gamma(c-a) \Gamma(c-b)}
F(a,b;a+b-c+1;1-x)  \\
& & 
+ \frac{\Gamma(c) \Gamma(a+b-c)}{\Gamma(a) \Gamma(b)} (1-x)^{c-a-b}
F(c-a,c-b;c-a-b+1;1-x) \, .\nonumber
\end{eqnarray}
We can introduce coefficients
\begin{eqnarray}
G_{i}^n (1-x;2 \leftrightarrow 3) &=& \sum_m c_{nm} G^m_{3-i} (x) \, ,
\end{eqnarray}
which are independent of the coefficient $i$, 
and obtain
\begin{eqnarray}
c_{11} &=& -\frac{\Gamma(2+e_1+e_3) \Gamma( -e_1-e_2-1)}{\Gamma(1+e_3) \Gamma(-e_2)}
\nonumber \\
c_{12} &=& -\frac{\Gamma(2+{e_1+e_3} ) \Gamma(1+ e_1+e_2 )}{\Gamma(1+{e_1} ) \Gamma(1+e_1+e_2+e_3 )}  \frac{1}{e_3}
\nonumber \\
c_{21} &=&   \frac{\Gamma(-e_1-e_3 ) \Gamma(-e_1-e_2-1 )}{\Gamma({-e_1} ) \Gamma(-e_1-e_2-e_3 )} 
e_2 
\nonumber \\
c_{22} &=& \frac{\Gamma(-e_1-e_3 ) \Gamma(1+ {e_1+e_2} )}{\Gamma(1-{e_3} ) \Gamma({e_2} )} \, .
\end{eqnarray}
 We require
\begin{eqnarray}
G(x,\bar{x}) &=& \sum_{i,j} X_{mm} G^m_{i}(x)  G^m_{j}(x) \tilde{I}_i \tilde{\bar{I}}_j
\nonumber \\
&=&  \sum_{i,j} X_{mm} G^m_{i}(1-x) G^m_j (1-x) (2 \leftrightarrow 3) \tilde{I}_{3-i} \tilde{\bar{I}}_{3-j}
\nonumber \\
&=& \sum_{i,j} X_{11}  (2 \leftrightarrow 3) c_{1m} G^m_{3-i}(x) c_{1n} G^n_{3-j}(x) 
 \\
 & & + X_{22}  (2 \leftrightarrow 3) c_{2m} G^m_{3-i} (1-x) c_{2n} G^n_{3-j}(1-x)) \tilde{I}_{3-i} \tilde{\bar{I}}_{3-j} \, .\nonumber
\end{eqnarray}
From these constraints, we find the equalities
\begin{eqnarray}
X_{11} (2 \leftrightarrow 3) c_{11} c_{12} + X_{22} (2 \leftrightarrow 3) c_{21} c_{22} &=& 0
\nonumber \\
X_{11} (2 \leftrightarrow 3) c_{11}^2 + X_{22} (2 \leftrightarrow 3) c_{21}^2 &=& X_{11}
\nonumber \\
X_{11} (2 \leftrightarrow 3) c_{12}^2 + X_{22} (2 \leftrightarrow 3) c_{22}^2 &=& X_{22} \, .
\end{eqnarray}
The first equation has the solution
\begin{eqnarray}
X_{11} &=&  - X_{22}
\Gamma(\frac{1+e_1,1-e_3,1+e_2,1+e_1+e_2+e_3,-e_1-e_2,-e_1-e_2}{
-e_1,e_3,2+e_1+e_2,2+e_1+e_2,-e_2,-e_1-e_2-e_3}) \, . \nonumber
\end{eqnarray}
This equation is identical to the constraint (\ref{localityconstraint}) we found from locality.
The functional dependence of the four point function is fixed by both requirements.
Crossing symmetry puts a further constraint on how the overall normalization behaves
under the exchange of the second and third insertion. We have to solve the equation
\begin{eqnarray}
X_{11} &=&  X_{11}(2 \leftrightarrow 3)
 \frac{\Gamma (-e_1-e_2-1)^2 \, \Gamma (2+e_1+e_3)^2}{  \Gamma (-e_2)^2 \, \Gamma (1+e_3)^2} \times
\nonumber \\
& & \csc(\pi e_2) \csc(\pi e_3) \sin (\pi(e_1+e_2)) \sin (\pi(e_1+e_3))\, .
\end{eqnarray}
One solution is
\begin{eqnarray}
X_{11} &=& \Gamma(\frac{-e_3,1+e_2}{2+e_1+e_2,-e_1-e_3-1})  \, .
\end{eqnarray}
We can multiply it by any function symmetric in the second and third momentum.
Our four point function satisfies both locality and crossing symmetry, for any value of the external momenta.

\subsubsection*{Remarks}
In \cite{Rozansky:1992rx}, crossing symmetry and locality of the correlator of four typical representations is analyzed
in the limit $e_2 + e_3 \rightarrow 0$. This limit gives rise to a logarithm in the chiral
block. In this limit, two typical representations tensor into a projective representation. The factorization of the 
four point function into three point functions in this regime was analyzed in \cite{Schomerus:2005bf}.

\subsection{The Local Projective and Three Typicals Correlator}
In this subsection, we construct local four point correlation functions for a top state in a projective
representation accompanied by three typical current algebra primaries. In particular, we will study the correlator
 $\langle t \uparrow \uparrow \downarrow \rangle$. We already determined that on the left it is of the form
\begin{eqnarray}
\langle t \uparrow \uparrow \downarrow \rangle_L &=& \frac{e_2+e_3}{e_3} G_1 + G_3
\end{eqnarray}
and we take it to have the same dependence on the right-movers, with $x$ replaced by $\bar{x}$, i.e. we choose the diagonal
vertex operator. The constants appearing
in the left chiral correlator become $\bar{x}$ dependent in the full correlator. The full local four-point correlation
function then satisfies both the left and the right Knizhnik-Zamolodchikov equations, and we need to glue the solutions into
a  local four-point function.

In order to find local four-point functions, we use again the technique of analyzing the chiral solutions near $x=0$ and near
$x=1$, to find combinations with phase monodromy. We can then take a linear combination of the modulus squared of these 
solutions to have a local solution near a given point. When we find a combination that is local at both $x=0$ and $x=1$,
we have a globally local solution. A small complication is that logarithms lead to shift monodromies, which we need
to cancel through recombination.

In our three step staircase solutions (\ref{3KZ}), we find the following behaviours near $x=0$ and near $x=1$
\begin{eqnarray}
\mbox{near} \, \, \,  x=0 \quad & : & \quad x^{e_2 n_1}, x^{e_2 (n_1+1)}, x^{e_2 n_1} \log x
\nonumber \\
\mbox{near} \, \, \, x=1 \quad & : &  \quad (1-x)^{e_3 n_1}, (1-x)^{e_3 (n_1+1)}, (1-x)^{e_3 n_1} \log (1-x) \, .
\end{eqnarray}
We have that the function multiplying $c_2$ corresponds to the $x^{e_2 (n_1+1)}$ behaviour, the function
multiplying $c_1$ to the 
$x^{e_2 n_1}$ limiting behaviour, and we tune the coefficients to 
\begin{eqnarray}
c_3 &=& c_1 \frac{ e_2  \csc \left(\pi  e_3\right) }{\left(e_2+e_3\right) \left(\pi  \csc \left(\pi  e_2\right) \csc \left(\pi  \left(e_2+e_3\right)\right)+\csc
   \left(\pi  e_3\right) \left(\psi \left(e_2\right)-\psi \left(e_2+e_3\right)\right)\right)}
\end{eqnarray}
and $c_2=0$ to obtain the logarithmic behaviour near $x=0$.
Near $x=1$, we similarly find the simple block multiplying $c_1$ with $(1-x)^{e_3 n_1}$ limiting behaviour, and the linear combinations obeying
\begin{eqnarray}
c_2 &=& c_1  \frac{ \sin \left(\pi  e_3\right) \Gamma \left(-e_3\right) \Gamma \left(e_2+e_3+1\right)}{\pi  \Gamma
   \left(e_2+1\right)} \, ,
\end{eqnarray}
with $c_3=0$ of $(1-x)^{e_3(n_1+1)}$ behaviour,
and
\begin{eqnarray}
\nonumber \\
c_3 & = & c_1  \frac{e_2}{\pi  e_2 \left(\cot \left(\pi  e_2\right)+\cot
   \left(\pi  e_3\right)\right)+ e_2 \psi \left(e_3\right) 
+ e_3 \psi(-e_2+1) - (e_2+e_3) \psi(-e_2-e_3+1)} 
\nonumber \\
c_2 &=& \frac{c_1}{\pi}
   \Gamma \left(1-e_2\right) \Gamma \left(-e_3\right) \Gamma \left(e_2+e_3+1\right) \nonumber \\
& & ( \pi \csc (\pi e_2) \csc (\pi e_3) e_2 + \csc (\pi (e_2+e_3)) \psi(e_3) e_2 
\nonumber \\
& & 
+ \csc(\pi (e_2+e_3)) \psi(-e_2+1) e_3 -
\csc \pi(e_2+e_3) \psi(-e_2-e_3+1) (e_2+e_3) )^{-1} \, , 
\end{eqnarray}
gives rise to a logarithmic limit.
We then define the blocks  with the above limiting behaviours (and $c_1$ equal to one, or $c_2=1$ when $c_1=0$), near $x=0$ and near $x=1$. We denote them
$B_{log}^{0,1},B_{n_1}^{0,1}, B_{n_1+1}^{0,1}$. To obtain four-point functions without monodromy, we define and equate
the combinations
\begin{eqnarray}
c^0_{log} (B_{log}^0  \bar{B}_{n_1}^0+ B_{n_1}^0 \bar{B}_{log}^0)
+c^0_{n_1} | B_{n_1}^0 |^2
+c^0_{n_1+1} |B_{n_1+1}^0 |^2
&=& \nonumber \\
 c^1_{log} (B_{log}^1  \bar{B}_{n_1}^1+ B_{n_1}^1 \bar{B}_{log}^1)
+c^1_{n_1} | B_{n_1}^1 |^2
+c^1_{n_1+1} |B_{n_1+1}^1 |^2 \, .
& & 
\end{eqnarray}
We find the constraints
\begin{eqnarray}
c^1_{log} &=& \frac{c^0_{log}\csc \left(\pi  e_3\right)  }{\left(e_2+e_3\right) \left(\pi  \csc \left(\pi  e_2\right) \csc \left(\pi 
   \left(e_2+e_3\right)\right)+\csc \left(\pi  e_3\right) \psi\left(e_2\right)-\csc \left(\pi  e_3\right) \psi
  \left(e_2+e_3\right)\right)}
\nonumber \\
& & 
\Big(e_3
   \left( \psi(-e_2+1)-\psi(-e_2-e_3+1) \right) \nonumber \\
& & +e_2 \left(-\psi(-e_2-e_3+1)+\pi  \cot \left(\pi  e_2\right)+\pi  \cot \left(\pi  e_3\right)+\psi
  \left(e_3\right) \right)\Big)
\nonumber \\
c^1_{n_1+1} &=&
\frac{\pi ^2 c^0_{n_1+1} \csc ^2\left(\pi  e_3\right) \Gamma \left(e_2+1\right){}^2}{\Gamma \left(-e_3\right){}^2 \Gamma
   \left(e_2+e_3+1\right){}^2}\nonumber \\
c^1_{n_1} &=& -c^1_{n_1+1}-2 c^1_{log}+2 c^0_{log} + c^0_{n_1}
\end{eqnarray}
which depend on our normalizations, as well as the final relation
\begin{eqnarray}
c^0_{log} &=& -\frac{\pi ^2 c^0_{n_1+1} \left(e_2+e_3\right) \csc \pi \left(e_2+e_3\right) \Gamma \left(e_2+1\right)
}{\Gamma \left(1-e_2\right) \Gamma \left(-e_3\right){}^2 \Gamma \left(e_2+e_3+1\right){}^2} 
\nonumber \\
& & 
\left(\pi  \csc
   \left(\pi  e_2\right) \csc \left(\pi  \left(e_2+e_3\right)\right)+\csc \left(\pi  e_3\right) \left(\psi \left(e_2\right)-\psi
   \left(e_2+e_3\right)\right)\right) \, .
\nonumber
\end{eqnarray}
The block $|B^0_{n_1}|^2$ is local at both $0$ and $1$, and we therefore find no constraint on the coefficient $c_{n_1}^0$
from locality. This term will also be crossing symmetric by itself. Otherwise, the full functional dependence of the four-point correlator is fixed by 
the requirement of locality.

 We remark that the four-point function continues the pattern
of the three-point function for one projective and two typical representations in that it is a combination of polygamma functions,
and more
particularly, harmonic numbers coded in digamma functions with given behaviour under integer shifts of the argument.

\section{Concluding Remarks}
\label{conclusions}
In this paper, we continued the study of the $gl(1|1)$ Wess-Zumino-Witten model. We
exploited that the quadratic Casimir acting on correlators is equal to
zero, and that the only non-zero one point functions are logarithmic partners
of group invariants.  The spectrum of the model consists of typical
and projective representations, and we explicitly exhibited 
correlation functions of projective representations in a covariant manner. We
provided a more general check of the crossing symmetry of typical
four point functions and computed a local projective and three typical
four point function, by solving a three step Knizhnik-Zamolodchikov
staircase.

Although the $gl(1|1)$ Wess-Zumino-Witten model is well-studied, there remain a slew of open questions. An open task is to compute
all three point functions and four point functions, and to show factorization on the proposed spectrum in full generality.
 It would also be instructive to map our Knizhnik-Zamolodchikov analysis 
onto the differential equations governing symplectic fermion correlators in detail, by gauging global symmetries.
In short, there is more groundlaying work to do on even the simplest of supergroup Wess-Zumino-Witten models.

This  task is important since what we learn in the most basic of logarithmic conformal field theories
with supergroup symmetry is bound to bare fruit in its diverse fields of application, including e.g.
 string theory on $AdS_3$ backgrounds, possibly with Ramond-Ramond flux.

\section*{Acknowledgments}
It is a pleasure to thank my colleagues  for providing a stimulating research environment. I
acknowledge  support from the grant ANR-13-BS05-0001.

\appendix

\section{The $gl(1|1)$ Algebra and  Representation Theory}
\label{reps}
We recall the part of the representation theory of the $gl(1|1)$ algebra that we need
in our analysis of the Wess-Zumino-Witten model.
The algebra is the algebra of endomorphisms of $\mathbb{C}^{1|1}$. These can be represented by  two by two matrices.
We can provide a basis $E_{ij}$ with $i,j=1,2$, and non-zero entry equal to 
one at the position $ij$. We define
\begin{eqnarray}
E = E_{11}+E_{22} \, , & \qquad &
N = \frac{1}{2}(E_{11}-E_{22})
\nonumber \\
\Psi^+ = E_{12} \, , & \qquad &
\Psi^- = E_{21} \, .
\end{eqnarray}
The $gl(1|1)$ algebra is $\mathbb{Z}$ graded, with two bosonic
generators $E,N$ of grade zero, and two fermionic generators 
$\Psi^\pm$ of grade plus and minus one respectively.
We recall the non-zero commutation relations of the $gl(1|1)$ algebra
\begin{eqnarray}
\{ \Psi^+, \Psi^- \} = E \, , 
& \qquad &
{[} N , \Psi^\pm {]} = \pm \Psi^\pm \, .
\end{eqnarray}
\subsection{The  Representations}
 We study representations in which
we diagonalize the actions of the central generator $E$ and the bosonic generator
$N$. (See e.g. \cite{Creutzig:2011np} for a summary.)  The
two-dimensional typical representation 
$T_{n-1/2,e}$ has a state with weights $(N,E)=(n,e)$ and one with eigenvalues $(n-1,e)$, where the eigenvalue $e$ is non-zero. We can represent the 
generators by the two by two matrices
\begin{eqnarray}
N = \left( \begin{array}{cc}
        n-1 & 0 \\
         0  & n 
       \end{array}  \right) \qquad
\Psi^+ = \left( \begin{array}{cc}
        0 & 0 \\
         e  & 0 
       \end{array}  \right)
\qquad
\Psi^- = \left( \begin{array}{cc}
        0 & 1 \\
         0  & 0 
       \end{array}  \right) \, .
\end{eqnarray}
The highest weight state can be a  boson or a fermion. We take the convention that for the constituent vertex operators in the typical
representations, the highest weight (up) state is a fermion
and the down state is a boson.\footnote{The tensor products representations inherit their parity from the constituting factors.}
We will  denote the two states through
\begin{eqnarray}
| \! \uparrow \rangle = \left( \begin{array}{c}
         0 \\
         1  
       \end{array}  \right)
\qquad
| \! \downarrow \rangle = \left( \begin{array}{c}
         1 \\
         0  
       \end{array}  \right)
\end{eqnarray}
such that
\begin{eqnarray}
\Psi^+ | \! \downarrow \rangle = e | \! \uparrow \rangle \qquad
\Psi^- | \! \uparrow \rangle = | \! \downarrow \rangle \, .
\end{eqnarray}
Other projective representations arise when $E$ has eigenvalue zero in the representation.
The projective cover is then a four-dimensional representation. The generators can
be represented by the four by four matrices:
\begin{eqnarray}
N = \left( \begin{array}{cccc}
         n & 0 & 0 & 0 \\
         0 & n+1 & 0 & 0   \\
         0 & 0 & n-1 & 0   \\
         0 & 0 & 0 & n  
       \end{array}  \right)
& \qquad & 
\Psi^+ = \left( \begin{array}{cccc}
         0 & 0 & 0 & 0 \\
         1 & 0 & 0 & 0   \\
         0 & 0 & 0 & 0   \\
         0 & 0 & 1 & 0  
       \end{array}  \right)
\nonumber \\
E=0
& \qquad &
\Psi^- = \left( \begin{array}{cccc}
         0 & 0 & 0 & 0 \\
         0 & 0 & 0 & 0   \\
         1 & 0 & 0 & 0   \\
         0 & -1 & 0 & 0  
       \end{array}  \right) \, .
\end{eqnarray}
One notation that we will use for the states is
\begin{eqnarray}
| \mbox{top} \rangle =  \left( \begin{array}{c}
         1 \\ 0 \\ 0 \\ 0  
       \end{array}  \right)
\qquad
| \mbox{left} \rangle =  \left( \begin{array}{c}
         0 \\ 1 \\ 0 \\ 0  
       \end{array}  \right)
\qquad
| \mbox{right} \rangle =  \left( \begin{array}{c}
         0 \\ 0 \\ 1 \\ 0  
       \end{array}  \right)
\qquad
| \mbox{bottom} \rangle =  \left( \begin{array}{c}
         0 \\ 0 \\ 0 \\ 1  
       \end{array}  \right)
\end{eqnarray}
such that we have
\begin{eqnarray}
\Psi^+ | \mbox{top} \rangle =  | \mbox{left} \rangle
& \qquad &
\Psi^- | \mbox{top} \rangle = | \mbox{right} \rangle
\nonumber \\
\Psi^+ | \mbox{right} \rangle = | \mbox{bottom} \rangle
& \qquad &
\Psi^- | \mbox{left} \rangle = - | \mbox{bottom} \rangle \, .
\end{eqnarray}
These projective representations are assembled from four atypical one-dimensional
representations of the algebra. In our convention, top states are bosons.
There is one state in these representations which is a trivial representation of the group.
It is the bottom state in the projective representation $P_0$. 
\subsection{Tensor Product Decompositions}
It will be important to
us to identify the occurrence of this invariant state (or, more precisely, its partner top state)
inside the tensor product of various typical and projective representations.
To find the projective representations $P_0$, it suffices to use the 
tensor product rules (see e.g. \cite{Creutzig:2011np})
\begin{eqnarray}
T_{n_1,e_1} \otimes T_{n_2,-e_1} &=& P_{n_1+n_2}
\nonumber \\
T_{n_1,e_1} \otimes T_{n_2,e_2} &=& T_{n_1+n_2+1/2,e_1+e_2} \oplus T_{n_1+n_2-1/2,e_1+e_2}
\nonumber \\
T_{n_1,e_1} \otimes P_{n_2} &=& T_{n_1+n_2+1,e_1} \oplus 2 T_{n_1+n_2,e_1} \oplus T_{n_1+n_2-1,e_1}
\nonumber \\
P_{n_1} \otimes P_{n_2} &=& P_{n_1+n_2+1} \oplus 2 P_{n_1+n_2} \oplus P_{n_1+n_2-1} \, .
\end{eqnarray}
\subsection{Counting Channels}
\label{channels}
We obtain the tensor product decompositions --  assuming that the total sum of the $E$-momenta $e_i$ is
equal to zero --:
\begin{eqnarray}
T_{n_1,e_1} \otimes T_{n_2,e_2} \otimes T_{n_3,e_3} & = &
P_{n_1+n_2+n_3+\frac{1}{2}} \oplus P_{n_1+n_2+n_3-\frac{1}{2}}
\nonumber \\
T_{n_1,e_1} \otimes T_{n_2,-e_1} \otimes P_{n_3} &=&
P_{n_1+n_2+n_3+1} \oplus 2 P_{n_1+n_2+n_3} \oplus P_{n_1+n_2+n_3-1} 
\nonumber \\
T_{n_1,e_1} \otimes T_{n_2,e_2} \otimes T_{n_3,e_3} \otimes T_{n_4,e_4} & = &
P_{\sum n_i +1} \oplus 2 P_{\sum n_i}  \oplus  P_{\sum n_i-1} 
\nonumber \\
P_{n_1}  \otimes T_{n_2,e_2} \otimes T_{n_3,e_3} \otimes T_{n_4,e_4}
 &=& 
P_{\sum n_i +3/2} \oplus 3 P_{\sum n_i+1/2}  \oplus 3 P_{\sum n_i-1/2} \oplus  P_{\sum n_i-3/2} \, .
\nonumber 
\end{eqnarray}
These give rise to the following counting of possible non-zero channels in correlation functions:
\begin{table}[H]
\centering
\begin{tabular}{|c|c|c|c|c|}
\hline
$\sum n_i$   & 0 & $\pm 1$ & $ \pm \frac{1}{2}$ & $ \pm \frac{3}{2}$
\\ 
\hline 
TT & 1 & - & - & - \\
PP & 2 & 1 & - & - \\
TTT & - & - & 1 & - \\
PTT & 2 & 1 & - & - \\
TTTT & 2 & 1 & - & - \\
PTTT & - & - & 3 & 1 \\
\hline
\end{tabular}
\caption{Counting projective $P_0$ representations in tensor products.}
\label{countingPzeroes}
\end{table}
\noindent
In appendix \ref{topandinvariants} we explicitly identify the invariant states in the projectives $P_0$
that appear in the tensor products of these representations.

\subsection{Invariant Combinations}
\label{topandinvariants}
In this subsection, we identify states in tensor products which are invariant under the action of the diagonal
algebra. These invariant states live at the bottom of projective $P_0$ representations.  We counted the
occurrences of the projective $P_0$ representations in table \ref{countingPzeroes}. We will exhibit 
invariant linear combinations of states below in each instance. These invariant combinations feature as  specific choices of basis in
the bulk of the paper.
\subsubsection{Two Typicals}
In the case of the tensor product of two typical representations satisfying $e_1+e_2=0$
and $n_1+n_2=0$, the invariant combination is (proportional to)
\begin{eqnarray}
I_{2,0;0} &=& \downarrow \uparrow - \uparrow \downarrow \, .
\end{eqnarray}
The lower indices indicate the number of typical representations in the tensor product, the number of 
projective covers $P$, and the negative of the sum of all quantum numbers $n_i$ labeling the representations.
\subsubsection{Two Projectives}
In the case of two projective $P_n$ representations, we distinguish three cases (and four invariants). We write down the invariant combinations
for the cases $n_1+n_2+1=0$ and $n_1+n_2=0$ (with the case $n_1+n_2-1=0$ mirroring what we find for $n_1+n_2+1=0$). For $n_1+n_2+1=0$,
we have
\begin{eqnarray}
I_{0,2;1} &=& b l - l b
\end{eqnarray}
and for $n_1+n_2=0$
\begin{eqnarray}
I^1_{0,2;0} &=& b b
\nonumber \\
I^2_{0,2;0} &=& bt+lr-rl+tb \, .
\end{eqnarray}
In this case, there are two invariants, and we have explicitly chosen a basis of invariants, where the upper label serves
as an enumeration of invariants.
\subsubsection{Three Typicals}
For the cases $\sum n_i  \pm 1/2=0$, we have the invariants
\begin{eqnarray}
I_{0,3;-1/2} &=& e_1 \uparrow \downarrow \downarrow + e_2 \downarrow \uparrow \downarrow + e_3  \downarrow  \downarrow \uparrow
\nonumber \\
I_{0,3;1/2} &=& 
\downarrow \uparrow \uparrow -
\uparrow \downarrow  \uparrow + \uparrow  \uparrow \downarrow
\, .
\end{eqnarray}
\subsubsection{One Projective and Two Typicals}
We put the projective representation in the first factor.
In the two channel $\sum n_i=0$ case, we have the invariants
\begin{eqnarray}
I^1_{2,1,0} &=& b \! \uparrow \downarrow + 
 l \! \downarrow \downarrow - 
e_2 (r \! \uparrow \uparrow + 
    t \! \downarrow \uparrow - 
    t \!  \uparrow \downarrow)
\nonumber \\
I^2_{2,1,0} &=& b \! \downarrow \uparrow -  b \! \uparrow \downarrow \, .
\end{eqnarray}
\subsubsection{Four Typicals}
\label{fourtypicalsinvariants}
Let us recall that we assume the total sum of the eigenvalues $e_i$ is zero. We have the cases $\sum n_i \pm 1=0$ with one invariant
and $\sum n_i =0$ with two.
If we concentrate on the invariant state with $\sum n_i+1=0$, we find that it is proportional to
\begin{eqnarray}
I_{4,0;1} &=&- \downarrow \uparrow \uparrow \uparrow +
   \uparrow \downarrow \uparrow \uparrow  -  \uparrow \uparrow \downarrow \uparrow +
  \uparrow \uparrow \uparrow \downarrow  \, .
\end{eqnarray}
For $\sum n_i=0$, we find two invariant channels. The invariant
linear combinations of states are
\begin{eqnarray}
I_{4,0;0}^1 &=&  e_3 ( \downarrow \downarrow \uparrow \uparrow  -
   \downarrow \uparrow \downarrow \uparrow  +    \downarrow \uparrow \uparrow \downarrow )
- e_1 (  \downarrow \uparrow \downarrow \uparrow  -
   \uparrow \downarrow  \downarrow \uparrow +     \uparrow \uparrow \downarrow \downarrow)
\nonumber \\
I_{4,0;0}^2 &=& -
 e_2 (  \downarrow \downarrow \uparrow \uparrow  -
    \downarrow \uparrow \downarrow \uparrow +    \downarrow \uparrow \uparrow \downarrow )
- e_1 (  \downarrow \downarrow  \uparrow \uparrow  -
    \uparrow \downarrow  \downarrow \uparrow  +     \uparrow  \downarrow \uparrow \downarrow ) \, .
\end{eqnarray}
\subsubsection{One Projective and Three Typicals}
 Let us consider the case in which we have
three possible channels, and where $\sum n_i +1/2=0$. We then solve for the three
invariant channels in this four factor tensor product space and find:
\begin{eqnarray}
I_{3,1;1/2}^1 &=& -\frac{e_3}{ e_2} 
(-b \! \uparrow \downarrow \uparrow + b \! \uparrow \uparrow \downarrow - l \! \downarrow \downarrow \uparrow + l \! \downarrow \uparrow \downarrow + e_2 
(r \! \uparrow \uparrow \uparrow + t \! \downarrow \uparrow \uparrow - t \! \uparrow \downarrow \uparrow + t \!
\uparrow \uparrow \downarrow)) 
\nonumber \\
I_{3,1;1/2}^2 &=& -e_3 (b \! \downarrow \uparrow \uparrow- b \! \uparrow \downarrow \uparrow + b \! \uparrow \uparrow \downarrow)
\nonumber \\
I_{3,1;1/2}^3 &=& \frac{1}{ e_2} (e_3(- b \! \uparrow \downarrow \uparrow + b \! \uparrow \uparrow \downarrow - l \! \downarrow \downarrow \uparrow + l \! \downarrow \uparrow \downarrow)
- e_2 ( b \! \uparrow \downarrow \uparrow + l \! \downarrow \downarrow \uparrow - l \! \uparrow \downarrow \downarrow)) \, .
\label{threechannels}
\end{eqnarray}
These parameterizations of channels are used in the bulk of the paper.

\subsection{The Tensor Products of Generators}
\label{Qs}
In this subsection, we compute the action of the tensor products of generators appearing in the Knizhnik-Zamolodchikov equation on the invariants
as parameterized in appendix \ref{topandinvariants}.
\subsubsection{Two Typicals}
The matrix coefficient $Q_{12}$ for acting on $I_{2,0;0}$ with the operator $Q_{12}$ is equal to $Q_{12}=-(h_{e_1,n_1}+h_{e_2,n_2})$, namely
the negative of the sum of conformal dimensions of the first and second insertion.
\subsubsection{Two Projectives}
For the $n_1+n_2+1$ channel, we find a zero coefficient, while for the $n_1+n_2=0$ channel, we find the matrix
\begin{eqnarray}
Q_{12} &=& \left( \begin{array}{cc}
0 & 0 \\
2  & 0 
\end{array}
\right)
\end{eqnarray}
acting on the basis of invariants $I_{0,2;0}^i$, namely the operator $Q_{1 \otimes 2}$ maps the second invariant into $2$ times the first.
\subsubsection{One Projective and Two Typicals}
In the two channel case we find the maps
\begin{eqnarray}
Q_{12} &=& \left(
\begin{array}{cc}
 e_2 n_1 & -e_2 \\
0 & e_2 n_1 \\
\end{array}
\right)
\nonumber \\
Q_{13} &=& \left(
\begin{array}{cc}
 -e_2 n_1 & -e_2 \\
 0 & -e_2 n_1 \\
\end{array}
\right) \, .
\end{eqnarray}
\subsubsection{Four Typicals}
In the one channel $\sum n_i+1=0$ case, we find that
$Q_{12}=h_{e_1+e_2,n_1+n_2+1/2} - h_{e_1,n_1}-h_{e_2,n_2}$.
In the two channel $\sum n_i=0$ case, we have
\begin{eqnarray}
Q_{12} &=&
\left(
\begin{array}{cc}
 \frac{e_2  \left(2 n_1+1 \right)+e_1 \left(2 e_2+2  n_2+1\right)}{2} & e_3 \\
 0 & \frac{e_2  \left(2 n_1-1\right)+e_1 \left(2 e_2+ 2 n_2-1\right)}{2} \\
\end{array}
\right)
\nonumber \\
Q_{13} &=& \left(
\begin{array}{cc}
 \frac{e_3  \left(2 n_1-1\right)+e_1 \left(2 e_3+ 2 n_3-1\right)}{2 } & 0 \\
 e_2 & \frac{e_3  \left(2 n_1+1\right)+e_1 \left(2 e_3+2  n_3+1\right)}{2 } \\
\end{array}
\right) \, .
\end{eqnarray}
\subsubsection{One Projective and Three Typicals}
The transformation matrices associated to the operators $Q_{1 \otimes 2}$ and $Q_{1 \otimes 3}$ are
\begin{eqnarray}
Q_{12} &=& \left(
\begin{array}{ccc}
 e_2 n_1 & 1 & 0 \\
 0 & e_2 n_1 & 0 \\
 0 & 0 & e_2 \left(n_1+1\right) \\
\end{array}
\right)
\nonumber \\
Q_{13} &=& \left(
\begin{array}{ccc}
 e_3 n_1 & -\frac{e_3}{e_2} & -e_3 \\
 0 & e_3 n_1 & 0 \\
 0 & \frac{e_3}{e_2}+1 & e_3 \left(n_1+1\right) \\
\end{array}
\right)\, .
\label{threechannelsQ}
\end{eqnarray}
We plug these matrices into the Knizhnik-Zamolodchikov equations in the bulk of the paper.

\section{Symplectic Fermions}
\label{symplecticfermions}
In this appendix, we remind the reader of results in \cite{Kausch:2000fu}, and complement them with some
extensions. First of all, we note that we have 
\begin{eqnarray}
L_0 \cdot : \chi^1 \chi^2: &=& 1
\end{eqnarray}
and we can thus set the rank 2 pair denoted $\omega,\Omega$ in \cite{Kausch:2000fu} equal to $\omega=:\chi^1 \chi^2$ and $\Omega =1$ in our framework.
We have the two point functions: \footnote{We fix the constant $O_\lambda$ in \cite{Kausch:2000fu} to $O_\lambda=1$.}
\begin{eqnarray}
\langle :\chi^1 \chi^2 : \rangle &=& 1
\nonumber \\
\langle :\chi^1 \chi^2 : :\chi^1 \chi^2 : \rangle &=&
-2(Z+\log |z_{12}|^2)
\nonumber \\
\langle \mu_{\lambda} \mu_{1-\lambda} \rangle &=& - |z_{12}|^{2 \lambda (1-\lambda)}
\end{eqnarray}
and the three point functions
\begin{eqnarray}
\langle  \mu_{\lambda} \mu_{1-\lambda} : \chi^1 \chi^2: \rangle &=&  |z_{12}|^{2 \lambda (1-\lambda)}
\left( Z_\lambda + \log | \frac{z_{13}z_{23}}{z_{12}}|^2 \right)
\nonumber \\
Z_{\lambda} &=& Z + 2 \psi(1)-\psi(\lambda)-\psi(1-\lambda)
\nonumber \\
\langle \mu_{\lambda_1} \mu_{\lambda_2} \mu_{\lambda_3} \rangle
&=& 
C_{\lambda_1,\lambda_2,\lambda_3} |z_{12}^{\lambda_1 \lambda_2}z_{13}^{\lambda_1 \lambda_3}z_{23}^{\lambda_2 \lambda_3}|^2
\qquad \mbox{for} \quad \lambda_1+\lambda_2+\lambda_3 =1
\nonumber \\
&=& C_{\lambda_1,\lambda_2,\lambda_3} |z_{12}^{(1-\lambda_1)(1- \lambda_2)}z_{13}^{(1-\lambda_1)(1- \lambda_3)}z_{23}^{(1-\lambda_2)(1- \lambda_3)}|^2
\quad \mbox{for} \, \,  \sum \lambda_i =2
\nonumber \\
C_{\lambda_1,\lambda_2,\lambda_3} &=& \sqrt{ \frac{\Gamma(\lambda_1)\Gamma(\lambda_2)\Gamma(\lambda_3)}{\Gamma(1-\lambda_1)\Gamma(1-\lambda_2)\Gamma(1-\lambda_3)}}
\end{eqnarray}
which we propose to supplement with the chiral expressions
\begin{eqnarray}
\langle \chi^1 \mu^L_{-\lambda+1} \mu^L_{\lambda+1} \rangle
&=& - \frac{1}{\lambda}  z_{12}^{\lambda} z_{13}^{-\lambda} z_{23}^{-\lambda^2}
\nonumber \\
\langle \chi^2 \mu^L_{-\lambda} \mu_\lambda^L \rangle
&=& -  z_{12}^{-\lambda} z_{13}^{\lambda} z_{23}^{-\lambda^2} \, .
\end{eqnarray}
We remark that the results on the four point function of the top primaries in the projective representations discussed in detail in the bulk
of the paper can 
be straightforwardly translated into new results for symplectic fermion correlators.

\bibliographystyle{JHEP}

\end{document}